\documentclass[11pt,a4paper]{article}
\usepackage{jcappub} 
\usepackage{comment}
\usepackage[utf8]{inputenc}
\usepackage{amstext}
\usepackage{amssymb}
\usepackage{amsmath}
\usepackage{amsfonts}
\usepackage{hyperref}
\hypersetup{
    colorlinks=true,
    linkcolor=blue,
    filecolor=magenta,
    urlcolor=cyan,
}
\usepackage[T1]{fontenc} 
\usepackage{breakcites}
\usepackage{graphicx}
\usepackage{float}
\usepackage{cleveref} 
\usepackage{subcaption}
\usepackage{placeins} 
\usepackage{color}
\usepackage{xcolor}
\usepackage[normalem]{ulem}  
\usepackage[numbers,sort&compress]{natbib} 
\usepackage{xspace}

\title{Stochastic gravitational wave background from stellar origin binary black holes  in LISA}
\date{November 2021}

\author[a]{Stanislav Babak,}
\author[b,c]{Chiara Caprini,}
\author[d]{Daniel G.~Figueroa,}
\author[e]{Nikolaos Karnesis,}
\author[f]{Paolo Marcoccia,}
\author[f]{Germano Nardini,}
\author[c,1]{Mauro Pieroni,%
\note{ Corresponding authors}} 
\author[g,l]{Angelo Ricciardone,}
\author[h,i,j]{Alberto Sesana,}
\author[l,m,1]{Jes\'us Torrado}

\affiliation[a]{Université de Paris, CNRS, Astroparticule et Cosmologie, 75013 Paris,
France}
\affiliation[b]{Université de Gen\`eve, Département de Physique Théorique and Centre for Astroparticle Physics, 24 quai
Ernest-Ansermet, CH-1211 Genéve 4, Switzerland}
\affiliation[c]{CERN, Theoretical Physics Department, 1 Esplanade des Particules, CH-1211 Genéve 23, Switzerland}
\affiliation[d]{Instituto de F\'isica Corpuscular (IFIC), Consejo Superior de Investigaciones Cient\'ificas (CSIC) and Universitat de Val\`{e}ncia, 46980, Valencia, Spain}
\affiliation[e]{Department of Physics, Aristotle University of Thessaloniki, Thessaloniki 54124, Greece}
\affiliation[f]{Department of Mathematics and Physics, University of Stavanger, NO-4036 Stavanger, Norway}
\affiliation[h]{Dipartimento di Fisica ``G. Occhialini'', Universit\`a degli Studi di Milano-Bicocca, Piazza della Scienza 3, I-20126 Milano, Italy}
\affiliation[i]{INFN, Sezione di Milano-Bicocca, Piazza della Scienza 3, I-20126 Milano, Italy}
\affiliation[g]{Dipartimento di Fisica ``E. Fermi'', Università di Pisa, I-56127 Pisa, Italy}
\affiliation[j]{INAF, Osservatorio Astronomico di Brera, Via E. Bianchi 46,  23807 Merate, Italy}
\affiliation[l]{Dipartimento di Fisica e Astronomia “G. Galilei”, Università degli Studi di Padova, via Marzolo 8, I--35131 Padova, Italy}
\affiliation[m]{INFN, Sezione di Padova, via Marzolo 8, I--35131 Padova, Italy}

\emailAdd{mauro.pieroni@cern.ch}
\emailAdd{jesus.torrado@pd.infn.it}

\abstract{We use the latest  constraints on the population of stellar origin binary black holes (SOBBH) from LIGO/Virgo/KAGRA (LVK) observations, to estimate the stochastic gravitational wave background (SGWB) they generate in the frequency band of LISA. 
In order to account for the faint and distant binaries, which contribute the most to the SGWB, we extend the merger rate at high redshift assuming that it tracks the star formation rate.
We adopt different methods to compute the SGWB signal: we perform an analytical evaluation, we use Monte Carlo sums over the SOBBH population realisations, and we account for the role of the detector by simulating LISA data and iteratively removing the resolvable signals until only the confusion noise is left.
The last method allows the extraction of both the expected SGWB and the number of resolvable SOBBHs. 
Since the latter are few for signal-to-noise ratio thresholds larger than five, we confirm that the spectral shape of the SGWB in the LISA band agrees with the analytical prediction of a single power law. 
We infer the probability distribution of the SGWB amplitude from the LVK GWTC-3 posterior of the binary population model: at the reference frequency of $0.003$ Hz it has an interquartile range of $\omhsqpivot \in [5.65,\,11.5]\times10^{-13}$, in agreement with most previous estimates. 
We then perform a MC analysis to assess LISA's capability to detect and characterise this signal.  
Accounting for both the instrumental noise and the galactic binaries foreground, with four years of data, LISA will be able to detect the SOBBH SGWB with percent accuracy, 
narrowing down the uncertainty on the amplitude by one order of magnitude with respect to the range of possible amplitudes inferred from the population model. A measurement of this signal by LISA will help to break the degeneracy among some of the population parameters, and provide interesting constraints, in particular on the redshift evolution of the SOBBH merger rate.}

\newcommand{\diff}{\mathrm{d}}
\newcommand{\sun}{\odot}
\newcommand{\yr}{\mathrm{yrs}}
\newcommand{\Runits}{\mathrm{Gpc}^{-3} \yr^{-1}}

\newcommand{\mmin}{{m_\mathrm{min}}}
\newcommand{\mmax}{{m_\mathrm{max}}}
\newcommand{\deltam}{{\delta_\mathrm{min}}}
\newcommand{\PP}{\textsc{Power Law\,+\,Peak}\xspace}
\newcommand{\FID}{\textsc{FidLVK}\xspace}
\newcommand{\pmint}[3]{{{#1}^{+#3}_{-#2}}}
\newcommand{\omhsqpivot}{h^2\Omega_\mathrm{GW}(f=3\times10^{-3}\,\mathrm{Hz})}
\newcommand{\taumax}{\tau_{c,\mathrm{max}}^{(\mathrm{det})}}

\begin{document}

\maketitle

\noindent

\section{Introduction}

Stellar-origin binary black holes (SOBBHs) are among the targets of the Laser Interferometer Space Antenna (LISA) \cite{2017arXiv170200786A}.
The emission of gravitational waves (GWs) from these binaries crosses the mHz frequency band, probed by LISA, while they are still far from coalescence.
Given the recent constraints from the LIGO/Virgo/KAGRA (LVK) collaboration after three observation runs, we expect a
large population of such systems contributing to the LISA data stream \cite{LIGOScientific:2021djp}.
At least a few of these binaries will be individually detected \cite{Sesana:2016ljz,2016MNRAS.462.2177K,2019PhRvD..99j3004G,Seto:2022xmh,Kremer:2018cir,Sesana:2019jmu}, while the bulk of them will form a stochastic GW background (SGWB), as they are too faint/distant and/or because they produce long-lived overlapping time-domain signals.
The characterization of both resolved and unresolved SOBBH sources is compelling since they are a source of confusion for other detectable sources in the LISA band.
For example, the SOBBH SGWB contribution will act as a foreground for the detection of a possible signal of cosmological origin~\cite{Caprini:2018mtu}, see e.g.~\cite{Lewicki:2021kmu,Boileau:2021,Boileau:2021kiw,Pieroni:2020rob} for prospects about the detectability of a cosmological SGWB in the presence of a SGWB of astrophysical origin.

Previous papers have estimated the expected level of the SOBBH background. 
This can be achieved via direct extrapolation of the LVK
observed merger rate, supplemented by a simple modelling of the black-hole (BH) population and of the time delay between the binary formation and the merger taken from \cite{Dominik:2012kk,2019MNRAS.490.3740N,Mapelli:2017hqk}, as done e.g.~in \cite{Sesana:2016ljz,Lewicki:2021kmu,Jenkins:2018uac,Mukherjee:2021ags}.
In particular, \cite{Lewicki:2021kmu,Mukherjee:2021ags}  use the LVK observed event rate from the Gravitational Wave Transient Catalog 2 (GWTC-2) \cite{LIGOScientific:2020kqk}.
Alternatively, one can input more refined scenarios of BH formation from the evolution of different populations of stars, accounting for the cosmic chemical evolution, optical depth to reionisation, and metallicity of the interstellar medium, to evaluate the mass distribution of merging SOBBH and in turn the expected SGWB, as done e.g.~in \cite{Dvorkin:2016wac,Nakazato:2016nkj,Perigois:2020ymr,Cusin:2019jpv,Cusin:2019jhg,Mangiagli:2019sxg}.
An estimate of the number of resolvable SOBBH in LISA using the GWTC-2 rate has been done e.g.~in \cite{Seto:2022xmh}.

In this paper, we employ several methods to estimate the SGWB in the LISA band, 
using the most recent population constraints from the Gravitational Wave Transient Catalog 3 (GWTC-3) \cite{LIGOScientific:2021djp}.
We evaluate the impact, on the SGWB amplitude, of the observational uncertainty on the population parameters, taken from the posterior parameter sample of GWTC-3: we find that the SGWB amplitude can vary by as much as a factor of five.
When considered independently, we show that the parameter whose marginalised $2\sigma$ error influences the most the SGWB level is the power-law index of the redshift dependence of the merger rate.
We also assess LISA’s capability to detect and characterise the predicted SOBBH SGWB via a Monte Carlo (MC) analysis of simulated data, including the SGWB, the galactic binary (GB) foreground component, and the instrumental noise.
The maximal marginalised error on the SGWB amplitude by LISA is $\sim5\%$, i.e.~much smaller than the variation due to the present (GWTC-3) observational uncertainty on the population parameters: this hints to the conclusion that LISA will have a role to play in constraining SOBBH population parameters via the SGWB measurement.
Though future Earth-based GW detectors observations will improve on the GWTC-3 constraints by the time LISA flies, we expect that LISA will maintain an impactful constraining power, since the SGWB amplitude in the LISA band 
is influenced by the high-redshift behaviour of the merger rate, complementary to what will be accessible to ground-based detectors in the near future. 

The paper is organised as follows.
In \Cref{sec:popmodandsynth} we describe the  population model that we use, and the assumptions we consider, to construct the SOBBH catalogues.
In particular, we disregard eccentricity in the waveform, as well as any redshift dependence of the population parameters, and we adopt a uniform distribution for the time-to-coalescence in the detector frame (\Cref{sec:popmodel}).
Faint and distant SOBBH contribute to the SGWB signal: we, therefore, need to complete the GWTC-3 merger rate, limited to low redshift, with a model for the star formation and evolution at higher redshift.
As explained in \Cref{sec:postparams}, we assume that the merger rate tracks the cosmic star formation rate (SFR) up to high redshift~\cite{Madau:2016jbv}.
We evaluate the impact of a time-delay between the binary star formation and the BBH merger on the SGWB amplitude in \Cref{sec:postpredictive}.
In \Cref{sec:popsynthesis} we describe the other population parameters: for the mass distribution we adopt the \PP model,
and for the spin amplitudes a positive-exponents Beta distribution \cite{LIGOScientific:2020kqk};
as for the remaining parameters, some of them are randomly generated (i.e.~time-to-coalescence, initial phase, position in the sky, inclination, and polarization), whereas others are derived analytically (e.g.~the initial frequency of the generated events, their distance...).
We have also produced ten SOBBH catalogues at a benchmark fixed point in the population parameter space, that we use for consistency studies; their characteristics are presented in \Cref{sec:fixedpoint}.

In \Cref{sec:backgroundmethods} we present the four methods we have used to  compute the expected SOBBH background signal.
In order of sophistication: (i) the first procedure is based on an analytic evaluation of the characteristic strain as an integral over the number density of inspirals, as first proposed by \cite{Phinney:2001di} (\Cref{sec:analytic});
(ii) we then substitute the integral over the number density by an MC sum over a realisation of a population, (iia) first as a time-to-coalescence-averaged sum, (iib) and then taking into account the time-to-coalescence of individual events and binning them according to their corresponding emission frequencies (\Cref{sec:montecarlo});
(iii) finally, in order to account for the actual detection process of the SOBBHs by LISA, we apply the iterative-subtraction method developed in \cite{Karnesis:2021tsh}, for which at each step we compare the signal-to-noise ratio (SNR) of each source $i$ ($\rho_i$) to an SNR threshold ($\rho_0$), and if $\rho_i > \rho_0$, the source is classified as resolvable and is subtracted from the data.
The iterative subtraction is performed on realistic LISA data-streams produced by injecting the time domain, spinning wave-form signals of the events, one by one. 
The latter procedure, despite being computationally expensive, yields a very accurate representation of the LISA data and allows for the evaluation of both the residual SGWB level and the subtracted sources (which we analyse in a companion paper \cite{next}).

In \Cref{sec:results} we present our results.
We first check that the four methods give comparable SGWB levels (\Cref{sec:comparison}): since the number of subtracted sources is small \cite{Perigois:2020ymr,next}, there is overall very good agreement.
Method (i), i.e.~the analytic integration of the background, while not capturing some detailed features of the signal, can safely be used to estimate the expected SGWB in the LISA band, for all points in the posterior parameter sample of the fiducial \FID model: the results are given in \Cref{sec:postpredictive}.
In \Cref{sec:backgroundparamsestimation} we present the results of the MC analysis of simulated LISA data including the SOBBH SGWB, the galactic binary foreground, and the instrumental noise.
We show that, also in presence of the GB foreground, with four years of data, LISA will be able to detect the SOBBH signal and reconstruct its amplitude and spectral index.
Then, accounting for the estimated SOBBH signal and the GB foreground as extra noise contributions, in \Cref{sec:powerlawsens} we build the LISA power-law sensitivity (PLS)~\cite{Thrane:2013oya, Caprini:2019egz, Flauger:2020qyi,Lewicki:2021kmu}.
Finally, in \Cref{sec:impactpopparams}, we analyse how the precise measurement of the SOBBH SGWB by LISA would impact the inference on population parameters, as put forward in \cite{Callister:2020arv,KAGRA:2021kbb}.
We find that the effect is most promising for the merger rate parameters, i.e.~amplitude, and power-law index. We conclude in \Cref{sec:conclusion}.

\section{SOBBH population model and use of GWTC-3 results}
\label{sec:popmodandsynth}

\subsection{SOBBH population model}
\label{sec:popmodel}

LISA is sensitive to the GW emission by SOBBHs in the inspiral phase.
Within the timescale of the mission, which we assume of 4 years (i.e.~4.5 years with 89\% duty cycle), the GW frequency emitted by most SOBBHs will slowly increase within the LISA frequency band, $f \in \left[ 10^{-4},0.1 \right] \mathrm{Hz}$.
A minority of SOBBHs will chirp (i.e.\ their GW emission will rapidly increase in frequency) and move throughout the band.
Among the chirping SOBBHs, a fraction will be close to coalescence, so that the frequency of their GW emission will exit the LISA band and, shortly after, enter the ground-based detectors band, where they will merge. 
This opens up the possibility of multi-band observations and/or of archival analyses (see e.g.~\cite{Sesana:2016ljz,Cutler:2019krq,Ewing:2020brd,Wong:2018uwb}).
On the other hand, no SOBBH entering the LISA band during the lifetime of the mission is statistically expected, as SOBBH with frequencies of the order of $10^{-4}$ Hz are practically monochromatic during the lifetime of the experiment.\footnote{To give an example, by integrating the Newtonian relation $\diff f_\mathrm{GW} / \diff t = 96/5 \, \pi^{8/3} (G\mathcal{M}/c^3)^{5/3} f_\mathrm{GW}^{11/3}$ 
one obtains that it takes about $10^8$ years to shift the GW emission of an SOBBH with chirp mass $\mathcal{M}=50 M_{\odot}$  from $2 \cdot 10^{-5}\ \mathrm{Hz}$ to $10^{-4}\ \mathrm{Hz}$, where it will still be about $10^6$ years away from the merger; 
while the same binary will shift from $0.1\ \mathrm{Hz}$ to $1\ \mathrm{Hz}$ (where it will be about 16 minutes away from the merger) in about 5 days.
}

We aim at estimating the SGWB due to unresolved SOBBHs in LISA, accounting for the most recent population constraints from GWTC-3 \cite{LIGOScientific:2021djp}.
For this aim, we generate catalogues of SOBBHs emitting in the LISA band, making some simplifying assumptions. 

First of all, for simplicity, we neglect eccentricity in our analysis. LVK measurements poorly constrain the SOBBH eccentricities, but the eccentricity in the LISA band could be significant depending on the binary formation \cite{2017MNRAS.465.4375N,Samsing:2018isx}.
In addition, we neglect
a possible dependence on redshift of the population parameters,
since there are no strong  constraints on how the SOBBH parameter distributions should vary with redshift, and state-of-the-art studies
based on observations have not found conclusive evidence on the presence of any redshift dependence  \cite{KAGRA:2021kbb,Mapelli_2019,LIGOScientific:2021psn} (this possibility has been explored e.g.\ in \cite{Karathanasis:2022rtr}).
Our methodology can incorporate a redshift dependence into the catalogue generation (albeit at a higher computational cost), if this will be constrained by future data.

Furthermore, we assume that the residual time to coalescence $\tau_c$ (i.e.\ the amount of time that an observer in the source frame must wait in order to see the binary merge) is statistically uniformly distributed across the SOBBH population.
This amounts to assuming that the formation, and therefore the coalescence rates, are in a steady state.
Indeed, any change in the demographics of the binaries happens on a cosmic time-scale of ${\cal O}(10^9)\,\yr$, i.e.~much longer than the LISA observation time, which is the typical time over which our catalogues are representative.
Furthermore, the maximal  $\tau_c$ that we consider in this analysis (c.f.~\Cref{sec:popsynthesis}) is $\tau_{c,\mathrm{max}}^{(\mathrm{det})}\sim {\cal O}(10^4)\,\yr$ in the detector frame, also much smaller than the timescale over which the cosmic coalescence rate varies.
We also neglect the possibility that the SOBBHs form on such a tight orbit that their GW emission at formation is already within the LISA band; this would indeed also break the uniform distribution hypothesis for $\tau_c$.

The above assumptions allow us to model the SOBBH population as follows.
We consider the binaries emitting in the LISA band and observed by the detector at a given instant
$t$, i.e.~the time at which LISA switches on.
Note that for the sake of the argument, we take this time in the source frame.
Among the intrinsic parameters (masses, spins, phase, polarisation\ldots) and extrinsic ones (sky position, inclination\ldots) of each SOBBH, we single out the  time-to-coalescence 
$\tau_c=t_c-t$ in the source frame (where $t_c$ denotes the time of coalescence of a given SOBBH) and the redshift of the source $z$, while $\vec{\xi}$ represents the remaining parameters.
The population model parameterized in terms of some hyper-parameters $\vec{\theta}$ provides the statistical distributions $p(\vec{\xi}|\vec{\theta})$ of $\vec{\xi}$ (for simplicity we omit the vector symbol on $\xi$ and $\theta$ from now on).
The number of SOBBHs
with given $z, \tau_c, \xi$, whose signals reach the interferometer at time $t$,  is
\begin{equation}
    \frac{\diff^3N(z, \tau_c, \xi,\theta)}{\diff\xi \diff z \diff\tau_c} = R(z, \tau_c)\left[\frac{\diff V_c}{\diff z}(z) \right] p(\xi | \theta)\, ,
    \label{eq:master}
\end{equation}
where $V_c$ is the universe's comoving volume,\footnote{The redshift derivative of the comoving volume in \Cref{eq:master} accounts for the fact that spherical shells further from us enclose increasing amounts of volume and thus larger numbers of events for a given $R(z)$.}
and
\begin{equation}
R(z,\tau_c) \equiv \frac{\diff^2N}{\diff V_c\, \diff\tau_c}\, .
\label{eq:R}
\end{equation}
Within our assumptions, all values of $\tau_c$ are equiprobable at any $z$: the rate density satisfies therefore $R(z,\tau_c)= R(z,\tau_c=0)$, i.e.~the one of {\it merging} SOBBHs.
Moreover, in our population, the number of SOBBHs with given $z$ and $\xi$ that are received by the interferometer at the times $t$ and $t+ dt$, are precisely
$N(z,\tau_c=0, \xi)$ and $N(z,\tau_c=0+dt, \xi)$, so one can equivalently interchange $ d\tau_c \leftrightarrow dt$ in \Cref{eq:R}.
All together, it follows $R(z,\tau_c)=\diff^2 N(z,\tau_c=0,\xi,\theta)/(\diff V_c\, \diff t)$, which is precisely the merger rate density in the form that LVK is constraining~\cite{LIGOScientific:2021psn}.
Hereafter, we drop the $\tau_c$ dependence in $R(z,\tau_c)$ as irrelevant.

Given our assumption that the merger rate $R(z)$ is in a steady state, we can readily apply LVK findings to it.
In the next section, we explain how we use \Cref{eq:master} to generate SOBBH catalogues compatible with the latest population constraints from LVK GWTC-3.
However, the merger-events-based LVK constraints on $R(z)$ are limited to small redshift, while we need to model sources also at high redshift, since they have a significant contribution to the SGWB.
In order to simulate the high-redshift part of the SOBBH population,
we therefore need to incorporate  knowledge of the star formation and evolution at high redshift, as we will see below.

\subsection{Implementing GWTC-3 posterior for the SOBBH population parameters}
\label{sec:postparams}

The SOBBH population model is determined by the merger rate $R(z)$ and the distribution function $p(\xi|\theta)$ in \Cref{eq:master}.
In \cite{LIGOScientific:2021psn}, the LVK collaboration has analysed a series of population models and produced inference on their parameters, finding that the most promising one to explain the SOBBH events gathered in  GWTC-3 \cite{LIGOScientific:2021djp} is characterised by: (a) a power-law dependence of the merger rate with redshift, $R(z)=R(0)(1 + z)^\kappa$; (b) a population mass model, known as \PP mass model, combining an inverse power-law dependence on the largest BH mass, with a Gaussian peak at approximately 30--40 $M_{\odot}$, and a power-law distribution for the mass ratio of the binary; (c) a population spin model in which the amplitudes are independent and follow positive-exponent Beta distributions favouring intermediate-valued spins, and whose tilt distribution is a mixture of an isotropic distribution and a truncated Gaussian. 
The distributions for masses and spins are explained in more detail in \Cref{app:denfun}.

For the sake of convenience, we will call this combination \FID, and we will use it as the fiducial model in our analysis.
We provide population-averaged predictions of the SGWB, based on the publicly-available population parameter posterior distribution \cite{ligo_scientific_collaboration_and_virgo_2021_5655785} for the \FID model conditioned to the SOBBH of GWTC-3 \cite{LIGOScientific:2021psn} (excluding low-mass-secondary GW190814 and likely-NSBHB GW190917, as per the fiducial approach by LVK).
The parameters $\theta$ of the mass and spin distribution $p(\xi|\theta)$ are imported directly from the LVK results.
On the other hand, the parameters of the merger rate $R(z)$  require a different treatment.
Because of the interferometers frequency band, LVK probe the SOBBH population only at relatively low redshift,
so the redshift dependence of the merger rate in LVK analyses is modelled as a power-law.
Indeed, the GWTC-3 inferred merger rate posterior
constrains the pivot rate $R(0)$ and the power-law exponent $\kappa$ only for $z\lesssim 0.5$ \cite{LIGOScientific:2021psn}.
In order to produce, from this posterior, SOBBH SGWB estimates valid in the LISA band, we need to extend the merger rate model towards higher redshift, since high redshift SOBBHs contribute significantly to the background.

For this purpose, we adopt a phenomenological approach and assume that the merger rate tracks the Madau-Fragos SFR \cite{Madau:2016jbv}, neglecting the presence of a time delay between the binary formation and merger.
While in \Cref{sec:postpredictive} we discuss the impact on the SGWB amplitude of including time delays, in the rest of the paper we parameterize the merger rate as
\begin{equation}\label{eq:rz}
    R(z) = R_{0} C \frac{(1 + z)^\kappa}{1 + \frac{\kappa}{r}\left(\frac{1 + z}{1 + z_\text{peak}}\right)^{\kappa + r}}\,,
\end{equation}
where $C$ ensures $R_0 \equiv R(z=0)$. 
The analysis of \cite{Madau:2016jbv} finds the following best fit values for the SFR parameters: $\kappa=2.6$, $r=3.6$ and $z_\text{peak}=2.04$ (note the difference in the definition of $z_\text{peak}$ with respect to \cite{KAGRA:2021kbb}). 
At redshift $z\lesssim 1$, this behaves similarly to the $R(z) = R_{0} (1 + z)^\kappa$ power law constrained by LVK, which finds a best fit $\kappa\approx 2.7$. Motivated by this agreement, we incorporate the LVK GWTC-3 posterior into \Cref{eq:rz} by matching $R_0$ and $\kappa$ for each point in the population parameter sample with the fixed fiducial values $r=3.6$ and $z_\text{peak}=2.04$ from \cite{Madau:2016jbv}. 
The resulting posterior for the merger rate is by construction fully compatible with that of the low-redshift merger rate of LVK (see \Cref{app:mergerrate} for further discussion and in particular \Cref{fig:merger_rate_comp}).\footnote{In~\cite{KAGRA:2021kbb}, the LVK Collaboration also considers a similar high-redshift extension and finds mild constraints on $r$ and $z_\text{peak}$ (using a different definition of the latter, see \Cref{app:mergerrate}) by combining the population parameters inferred from GWTC-2
resolved mergers with the upper limits imposed by the non-detection of the SGWB.
We verify a posteriori the compatibility of our results with the upper limits on the SGWB amplitude presented in~\cite{KAGRA:2021kbb}.}

Finally, in order to keep consistency with LVK \cite{LIGOScientific:2021psn}, we adopt the $\Lambda$CDM cosmological model with parameters fixed accordingly to the ``Planck 2015 + external'' data combination  \cite{Planck:2015fie}.\footnote{Note that the $\Lambda$CDM parameter values used in \cite{LIGOScientific:2021psn} correspond to the incomplete Planck 2015 data combination \textsc{plikHM\_TE} (high-$\ell$ T$\times$E spectrum data only) instead of the fiducial \textsc{plikHM\_TTTEEE+lowTEB}, which includes temperature-only and polarized data for the whole Planck multipole range. The difference is anyway negligible for the purposes of this paper.}
These correspond to $H_0=67.9\,\mathrm{km}/(\mathrm{s\,Mpc})$ for the local Hubble rate, and $\Omega_m\approx 0.3$ and $\Omega_\Lambda\approx 0.7$ for the matter and cosmological-constant energy densities.
The cosmological model enters in the differential comoving volume per unit redshift, $\diff V_c(z)/  \diff z$, of \Cref{eq:master}, and in the computation of the cosmological distances needed for the
integration in \Cref{sec:backgroundmethods}.

\subsection{SOBBH population synthesis}
\label{sec:popsynthesis}

In \Cref{sec:backgroundmethods} we propose four different methods to compute the SGWB due to unresolved SOBBHs. 
The first method consists of an integration of the number density in \Cref{eq:master}
\cite{Phinney:2001di}. 
The other three are based on the superposition of the GW signals from SOBBHs populations, with different levels of sophistication.
The latter methods provide a more refined evaluation of the SGWB and of its spectral shape and are 
also important to assess the size of statistical effects (e.g.~the uncertainty due to the population realisation) and  
the consequences of other choices inherent to the catalogue simulation, such as the value of the maximal time-to-coalescence 
$\taumax$, see below. 

We thus need fast and reliable SOBBH population synthesis. 
We have written two independent population synthesis codes, which can be found in the following repositories: \cite{SynthesisCodePaolo} and \cite{SynthesisCodeJesus}. These have been also compared with \cite{Pieroni:2022bbh}.
In these implementations, the masses and spins are drawn from the LVK GWTC-3 distributions, briefly revised in \Cref{app:denfun}.

The redshift of the binaries is generated independently as an inhomogeneous Poisson point process,
according to the $z$-dependent terms in \Cref{eq:master},
between $z_\mathrm{min}=10^{-5}$ ($\approx 45\,\mathrm{kpc}$ of comoving distance, in order to exclude binaries within the Milky Way), and $z_\mathrm{max}=5$, which is sufficient for an accurate SGWB computation, as we demonstrate in \Cref{sec:analytic}.
Note that we will limit $z_\mathrm{max}=1$ 
in the analyses based on catalogues whenever using a larger $z_{\rm max}$ would prove too costly from the computational point of view, c.f.~\Cref{sec:iterative}.

The rest of the individual SOBBH parameters are generated from the priors presented in \Cref{tab:USParametersPrior}, based on physical considerations: isotropy for the sky position, inclination, and polarization; and uniform time-to-coalescence in the detector frame, as discussed in \Cref{sec:popmodel}. 
From the randomly-sampled parameters, we compute the derived quantities necessary for the problem at hand, such as the frequency at the start of the LISA runtime, the LISA in-band time, cosmological distances, and so on. 

\begin{table}[t]
\centering
\begin{tabular}{cc}
\textbf{Parameter} & \textbf{Prior} \\ \hline\hline
Time-to-coalescence (source frame) & $U[0, \tau_{c,\mathrm{max}}^{(\mathrm{det})}/(1+z)] \; \yr$ \\
Ecliptic Longitude & $U[0,2 \pi] \; \mathrm{rad}$ \\
Ecliptic Latitude & $\arcsin\left(U[-1,1]\right) \; \mathrm{rad}$ \\
Inclination & $\arccos\left(U[-1,1]\right) \; \mathrm{rad}$ \\
Polarization & $U[0,2 \pi] \; \mathrm{rad}$ \\
Initial Phase & $U[0,2 \pi] \; \mathrm{rad}$\\
\hline
\end{tabular}
\caption{Priors for the parameters of individual SOBBHs.
The uniform prior for the time-to-coalescence, which is source-dependent, is justified in \Cref{sec:popmodel}. The priors on the ecliptic coordinates and the inclination impose statistical isotropy in the positions and orientation of the binaries.}
\label{tab:USParametersPrior}
\end{table}

The upper limit for the population synthesis time-to-coalescence $\tau_{c,\mathrm{max}}^{(\mathrm{det})}$ needs to be high enough to give a faithful representation of the SOBBH SGWB signal in the LISA band (at least where it is the dominant contribution to the astrophysical-origin SGWBs), and at the same time it is conditioned by computational limitations. 
As discussed in \Cref{app:ttot}, $\tau_{c,\mathrm{max}}^{(\mathrm{det})}=10^4\,\yr$ provides a good balance between these requirements.

\subsection{Benchmark fixed-point catalogues for consistency studies}
\label{sec:fixedpoint}

In addition to probabilistic GWTC-3-posterior forecasts, we also single out a fixed point in the population parameter space, which we use as a benchmark to compare different SGWB computation methods and assess the size of statistical and numerical effects. 
For this fixed point, we use values close to the median of the GWTC-3 \FID model posterior, indicated in \Cref{tab:N18params}, with an important modification.

\begin{table}[ht]
\centering
\begin{tabular}{c|c|c}
\textbf{Rate of events} $R(z)$ & \textbf{Mass distribution} & \textbf{Spin distribution} \\\hline\hline
  \begin{tabular}{ll}
    $R_{0.2}=28.1\,\Runits$\\
    $\kappa=2.7$\\
    $z_\mathrm{peak}=2.04$\\
    $r=3.6$ \\
    \\
    \\
  \end{tabular}
&
  \begin{tabular}{ll}
    $[\mmin, \mmax]\,\in [2.5, 100]\,M_\sun$ \\
    $\deltam =7.8\,M_\sun$ \\
    $\alpha=3.4$ \\
    $\lambda_\mathrm{peak}=0.039$\\
    $\mu_m=34\,M_\sun$ \\
    $\sigma_m=5.1\,M_\sun$ \\
    $\beta_q=1.1$ \\
  \end{tabular}
&
  \begin{tabular}{ll}
    $\mathrm{E}[a]=0.25$ \\
    $\mathrm{Var}[a]=0.03$ \\
    $\zeta=0.66$ \\
    $\sigma_t=1.5$ \\
    \\
    \\
  \end{tabular}
\end{tabular}
\caption{Population parameter values for the benchmark fixed-point. The other parameters in the population are set to their GWTC-3 posterior median values, given in \Cref{app:mergerrate,app:denfun}. 
The mass range and mass smoothing parameter $\deltam$ have been modified to accommodate for the possibility of more extreme events in future data. 
}
\label{tab:N18params}
\end{table}

The determination of the mass range population parameters $\mmin$ and $\mmax$ in the LVK study~\cite{LIGOScientific:2021psn} is sensitive to whether certain events from GWTC-3 (the extreme mass ratio binary GW190814 and the likely-NSBHB binary GW190917) are considered as 
outliers and excluded from the analysis. The inclusion of GW190814 suffices to push the lower mass bound down to $\mmin\approx 2.5\,M_\sun$.

Motivated by the 
possibility that such outliers may appear in future data, we enlarge the mass range for the benchmark fixed-point catalogues to $[2.5, 100]\,M_\sun$, also raising the upper bound up to the original prior boundary of the $\mmax$ population parameter, previous to GTWC-3 constraints. 
The modification in particular of the lower mass boundary necessitates a further increase in the width of the low-mass smoothing function (see \Cref{app:denfun}), in order not to deviate strongly from the mean GWTC-3 mass probability density at masses $m\gg \mmin$. This is achieved by increasing the value of the $\deltam$ population parameter (nevertheless, we have found that the SGWB calculation is not very sensitive to this choice).

Though this fixed point in the population parameter space 
does not pertain to the GWTC-3 posterior, due to the modifications to the mass distribution, 
it leads to an SGWB in the LISA band which is compatible with our posterior-based evaluations. 
It is therefore useful as a benchmark to gauge the 
sensitivity of the SGWB predictions to different assumptions on the population model. 

We have generated a sample of 10 catalogues with parameters set to this benchmark fixed-point. 
Since they will be used only to compare 
and validate population-based SGWB computation methods, we have limited their redshift range to $z_\mathrm{max} = 1$, to reduce computational cost. 
As stated before, we limit the time-to-coalescence in these catalogues to $\tau_{c, \mathrm{max}}^{(\mathrm{det})}=10^4\,\yr$. 
For the purposes of testing the sensitivity to different $\tau_{c, \mathrm{max}}^{(\mathrm{det})}$ values  (see \Cref{sec:iterative}), 
we have generated one catalogue with $\tau_{c, \mathrm{max}}^{(\mathrm{det})}=1.5\times10^4\,\yr$, and from it we have produced two sub-catalogues with $\tau_{c, \mathrm{max}}^{(\mathrm{det})}=1.0\times10^4$ and $5\times10^3\,\yr$. 
Setting $\tau_{c,\mathrm{max}}^{(\mathrm{det})}=10^4\,\yr$ produces approximately 60 million binaries with inspiralling frequency within the LISA band. 
The number of events scales linearly with $\tau_{c,\mathrm{max}}^{(\mathrm{det})}$.

\section{Computation of the SOBBH signal in the LISA band }
\label{sec:backgroundmethods}
We adopt four different methods to evaluate the SOBBH SGWB which allow us, by their different nature, to capture different features of the signal. In the following sections, we describe them.

\subsection{Method (i): analytical evaluation}
\label{sec:analytic}

In this section we provide a brief description of the formalism employed for the analytic evaluation of the SOBBH SGWB, following \cite{Phinney:2001di}. 
The normalised SGWB energy density spectrum per logarithmic unit of frequency $\Omega_\mathrm{GW}(f)$ can be defined from the total GW energy density present in the universe and emitted by the whole SOBBH population, expressed in the detector frame. 
Recalling \Cref{eq:master} this reads
\begin{equation}
\label{eq:omega}
\frac{\rho_\mathrm{GW}^{\rm (tot)}}{\rho_c}=\int_0^\infty \frac{\diff f}{f}\,\Omega_\mathrm{GW}(f)  = \int \diff \xi \int \diff V_c \int \diff \tau_c \, \frac{\diff^3N(z, \tau_c, \xi,\theta)}{\diff\xi \diff V_c \diff \tau_c} \,\frac{\rho_\mathrm{GW}^{\rm (event)}}{\rho_c}\,,
\end{equation}
where $\rho_c =3 H_0^2 c^2/(8 \pi G)$ is the Universe's critical energy density and $\rho_\mathrm{GW}^{\rm (event)}=t^{00}$ denotes the energy density associated to a single SOBBH event, at the detector.
Using \cite{Maggiore:2007ulw}, one can derive
\begin{align}
    \lefteqn{\frac{\rho_\mathrm{GW}^{\rm (event)}}{\rho_c}= \frac{c^2}{16\pi G \rho_c}\frac{\langle\dot h_+^2+\dot h_\times^2\rangle}{(1+z)^4}=} \label{eq:rhoGWsingle} \\
     & & = \frac{c^2}{16\pi G \rho_c (1+z)^4}
    \frac{32\pi^2}{a^2 r^2}
    \Big(\frac{\pi}{c}\Big)^{4/3}   
    \Big(\frac{G \mathcal{M}}{c^2}\Big)^{10/3}
    f_s(\tau_c)^{10/3}
    \Big[\Big(\frac{1+\cos^2 \iota}{2}\Big)^2+\cos^2\iota\Big]\,, \nonumber
\end{align}
where all quantities are at the source: $\mathcal{M}$ is the SOBBH chirp mass, $a\!\cdot\!r$ its physical distance in the local wave zone, $\iota$ its orientation with respect to the detector, and $\omega_s=\pi f_s$ its orbital frequency. 
The second equality in \Cref{eq:rhoGWsingle} has been obtained under the approximation of quasi circular motion for the binary $\dot f_s\ll f_s^2$, and we have averaged over the waveform phase.  
Substituting \Cref{eq:rhoGWsingle} in \Cref{eq:omega}, expressing the differential comoving volume as $\diff V_c=c\,d_M^2/H(z) \diff \hat\Omega \diff z$ where $d_M=a_0 r$ is the metric distance,  $ H(z) = H_0 \sqrt{\Omega_m (1+z)^3 + \Omega_{\Lambda}}$, and $\hat\Omega$ is the solid angle \cite{Hogg:1999ad}, and noting that $a^2 r^2=d_M^2/(1+z)^2$, one gets, after integration over the solid angle (giving a factor $16\pi/5$),
\begin{equation}
    \frac{\rho_\mathrm{GW}^{\rm (tot)}}{\rho_c}  
    = \frac{32}{5}\pi^{10/3}\frac{c^{5/3}}{G\,\rho_c}
    \int \diff \xi \int \diff z \int \diff \tau_c 
    \Big(\frac{G \mathcal{M}}{c^2}\Big)^{10/3}
   \,\frac{R(z)p(\xi|\theta)}{H(z)(1+z)^2}  f_s(\tau_c)^{10/3}\,,
    \label{eq:omega_interm}
\end{equation}
where we have also used definition \eqref{eq:master}.
One can change the integration variable from $\tau_c$ to $f_s$ using the relation $\diff f_s /\diff \tau_c= 96/5 \, \pi^{8/3} (G\mathcal{M}/c^3)^{5/3} f_s^{11/3}$, valid for quasi circular binaries in the Newtonian approximation, then change to the frequency at the detector
$f=f_s/(1+z)$, and equate the integrands in \Cref{eq:omega}, to obtain the SGWB energy density power spectrum:
\begin{equation}
    \Omega_\mathrm{GW}(f)=\frac{8}{9}\pi^{5/3}f^{2/3}
    \int \diff \xi \int \diff z \frac{(G \mathcal{M})^{5/3}}{c^2 H_0^2}
   \,\frac{R(z)p(\xi|\theta)}{H(z)(1+z)^{4/3}}\,.
   \label{eq:Omega_int}
\end{equation}
Among the set of binary parameters $\xi$, only the chirp mass is relevant within the Newtonian approximation. 
One can therefore express the SOBBH SGWB today as\begin{equation}\label{eq:powerlaw}
h^2\Omega_\mathrm{GW}(f) = h^2\Omega_\mathrm{GW}(f_*)\left(\frac{f}{f_*}\right)^{2/3}\,,
\end{equation}
with $f_*$ an arbitrary pivot frequency, and
\begin{equation}
\label{eq:CoeffOmegaGWAnalytic}
\begin{split}
h^2\Omega_\mathrm{GW}(f_*) = &\, \frac{(GM_{\sun})^{5/3}}{c^2}
     \int \diff m_1 \diff m_2\,p\left(m_1,m_2\right)(\mathcal{M}(m_1,m_2)[M_{\sun}])^{5/3}\\
     & \times
     \int_{0}^{z_\mathrm{max}} \diff z \, \frac{R\left(z\right)}{(1+z)^{4/3}H(z)} \frac{f_*^{2/3}} {(1.32413\times 10^{-18}\, {\rm Hz})^2}\,,
\end{split}
\end{equation}
where we have used $(H_0/h)/\sqrt{8\pi^{5/3}/9}  = 1.32413\times 10^{-18}$ Hz. 
For the numerical evaluation of \Cref{eq:CoeffOmegaGWAnalytic}, we have set $GM_{\sun} = 1.327\times 10^{20}$ m$^3$/s$^2$ and $c = 2.9979246 \times 10^8$ m/s.

\begin{figure}[t]
    \includegraphics[width=1.0\textwidth,angle=00 ]{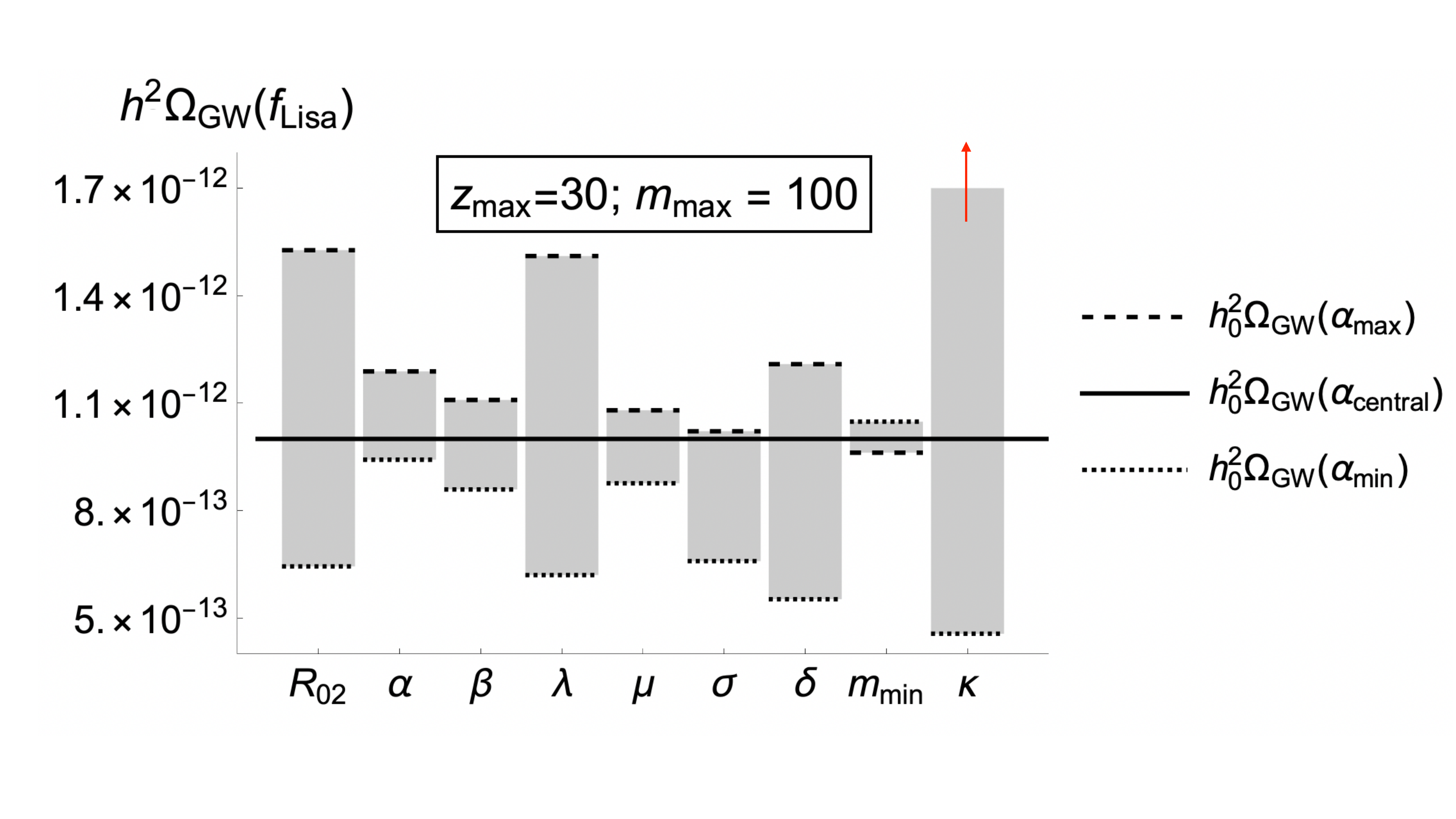}
    \vspace*{-0.75cm}
  \caption{Amplitude $h^2\Omega_\mathrm{GW}(f_*)$ of the SOBBH SGWB at the pivot frequency $f_* = f_\mathrm{LISA} \equiv 0.003$ Hz, varying each parameter of the \FID population model within the $5$--$95$ percentile range of the GWTC-3 posterior, while maintaining all other parameters fixed to their central values. Dotted (dashed) lines represent the value of $h^2\Omega_\mathrm{GW}(f_*)$ evaluated at the $5$ ($95$) percentile for a given parameter, whereas the horizontal solid line represents the amplitude of the signal when all parameters are fixed to the values in \Cref{tab:N18params}.
  The red arrow in $\kappa$ indicates that the amplitude of the signal for $\kappa_\mathrm{max}$ ($h^2\Omega_\mathrm{GW}(f_*) \simeq 2.8\times 10^{-12}$) is beyond the range of the plot. The signal amplitude grows with decreasing $m_{\rm min}$, contrary to all other parameters.
  }
\label{fig:analyticalSignal}
\end{figure}

\begin{figure}[t]
     \centering
    \includegraphics[width=0.65\textwidth,angle=00]{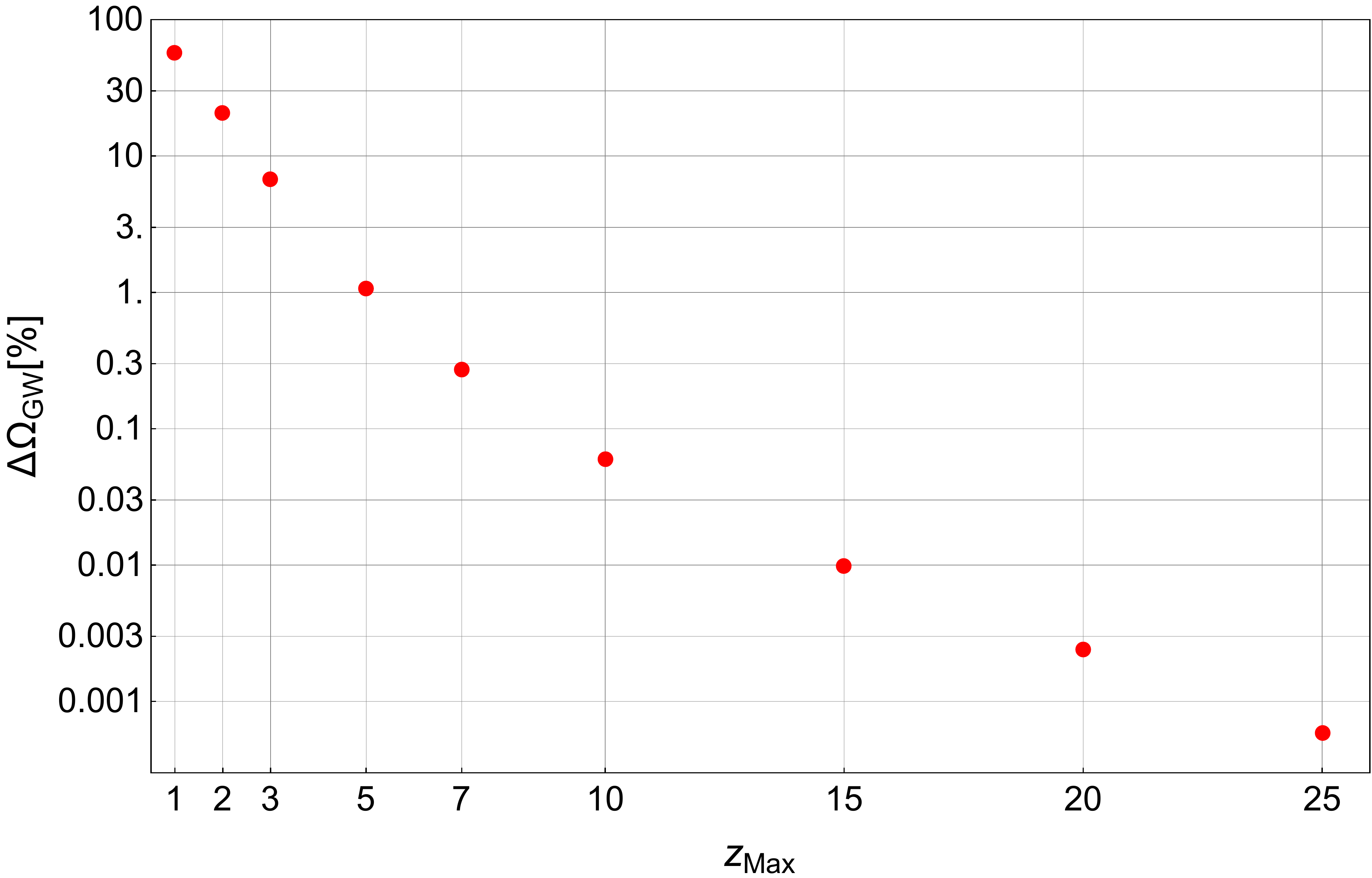}~~
  \caption{Relative percentage change of the signal amplitude $\Omega_\mathrm{GW}(f_*)$ evaluated at $f_* = 0.003\,\mathrm{Hz}$ with respect to $\Omega_\mathrm{GW}(f_*)|_{z_\mathrm{max}=30}$, when varying $z_\mathrm{max}$ in \Cref{eq:CoeffOmegaGWAnalytic}.}
\label{fig:analyticalSignalII}
\end{figure}

In \Cref{fig:analyticalSignal} we plot the amplitude of the expected background at the reference frequency $f_* = 0.003\,\mathrm{Hz}$, close to LISA's peak sensitivity \cite{Babak2021mhe}, evaluated from the integral in Eq.\ \eqref{eq:CoeffOmegaGWAnalytic}. 
For the merger rate $R(z)$, we adopt the phenomenological parameterization described in \Cref{sec:postparams}. While in \Cref{fig:analyticalSignal} we consider $z_\mathrm{max}=30$, as we do not expect active sources at higher redshifts, in~\Cref{fig:analyticalSignalII}, in contrast, we analyse the relative difference of considering smaller values for $z_\mathrm{max}$.
The only other population parameters that enter the analytic evaluation are the masses $m_1, m_2$ of the two compact objects, expressed in terms of the chirp mass: as previously stated, for their probability distribution $p\left(m_1,m_2\right)$, we adopt the \PP model.
Naturally, the amplitude of the background depends on the choice of the parameters in $R(z)$ and $p(m_1,m_2)$: respectively, $(R_{02}\equiv R(z=0.2),\kappa)$, and $(\alpha, \deltam, \mmin, \mmax, \lambda_\mathrm{peak}, \mu, \sigma, \beta)$, defined in \Cref{app:mergerrate,app:denfun}. 
Following \Cref{sec:fixedpoint}, we plot as a horizontal solid line the SGWB amplitude for the values indicated in \Cref{tab:N18params} for each of the population parameters
; we also show the range of SGWB amplitudes (grey bars) obtained when each of the parameters is varied within its $5$--$95$ percentile range according to the GWTC-3 posterior (see \Cref{app:mergerrate,app:denfun}), while the rest of the parameters stay fixed to their values of \Cref{tab:N18params}. 
This shows how the different parameters in the model influence the SGWB amplitude when varied individually. 
Larger ranges for the SGWB amplitude translate into stronger constraining power from the measurement of the individual parameter; 
however, since this approach neglects degeneracies, large ranges for multiple parameters do not mean that these can be simultaneously constrained (the issue of SGWB-derived population parameter constraints will be further discussed in \Cref{sec:impactpopparams}). 
Note that the signal amplitude grows with decreasing $m_{\rm min}$, contrary to its response to all other parameters. 
The population parameter with the largest impact on the SGWB amplitude is the power-law index of the merger rate $\kappa$, since its value controls the 
merger rate growth at intermediate redshift $1<z < z_{\rm peak}$, which strongly influences the outcome of the redshift integration in \Cref{eq:CoeffOmegaGWAnalytic}.
The red arrow in \Cref{fig:analyticalSignal} indicates that the SGWB amplitude obtained for $\kappa_\mathrm{max}$, $h^2\Omega_\mathrm{GW}(f_*) \simeq 2.8\times 10^{-12}$, is beyond the range of the plot.

In \Cref{fig:analyticalSignalII} we plot the relative percentage change of the SGWB amplitude when varying $z_\mathrm{max}$ in \Cref{eq:CoeffOmegaGWAnalytic}. 
Since the merger rate in \Cref{eq:rz} decays at high redshift, the SGWB grows asymptotically towards a constant amplitude as we integrate over larger and larger redshift ranges. 
Taking $z_\mathrm{max} = 30$ as a reference, we plot $\Delta\Omega_\mathrm{GW}[\%] \equiv 100\times\left(1-{\Omega_\mathrm{GW}(f_*)|_{z_\mathrm{max}}}/{\Omega_\mathrm{GW}(f_*)|_{z_\mathrm{max}=30}}\right)$ for different values of $z_{\max}$.
The figure indicates that integrating up to $z_\mathrm{max} = 5$ already allows to obtain $\sim 1\%$ accuracy in the calculation of the SGWB amplitude. 
This is sufficient for the  scope of this paper given that, as presented in \Cref{sec:backgroundparamsestimation}, the typical error on the SGWB measurement by LISA is larger than that. 
The SGWB amplitude in \Cref{fig:analyticalSignalII} has been evaluated adopting the parameter values corresponding to the benchmark fixed-point described in \Cref{sec:fixedpoint}; the convergence trend is very similar when using the minimum or maximum values of a given parameter, keeping the others fixed to their central values.

\subsection{Methods (iia) and (iib): Monte Carlo sum}
\label{sec:montecarlo}

An alternative method to compute the SGWB is to sum the  GW signals, emitted by individual SOBBHs, over a realisation of the population drawn from the distribution represented by the number density.
The simplest implementation consists in factoring out from  \Cref{eq:Omega_int} the population number density of \Cref{eq:master}, averaged over time-to-coalescence and sky-position, to obtain
\begin{equation}\label{eq:char_strain_mcsum}
  \Omega_{\rm GW}(f) \approx \frac{2\pi^{2/3}}{9} \frac{G^{5/3}}{c^3 H_0^2}\frac{1}{\tau_{c,\mathrm{max}}^\text{(det)}} \left(
    \sum_{i\in\text{pop}} \frac{\mathcal{M}_i^{5/3}}{d_{M,i}^2 (1+z_i)^{1/3}}
  \right) f^{2/3}\,,
\end{equation}
where $f$ is the observed frequency, and $\mathcal{M}_i$, $z_i$ and $d_{M, i}$ are the chirp mass in the source frame, redshift and metric distance of the individual GW sources. 
The factor $1/\tau_{c,\mathrm{max}} = (1+z)/\tau_{c,\mathrm{max}}^\text{(det)}$ comes from the time-averaging of the number density of \Cref{eq:master}.

\begin{figure}[t!]
    \centering
    \includegraphics[width=0.75\textwidth]{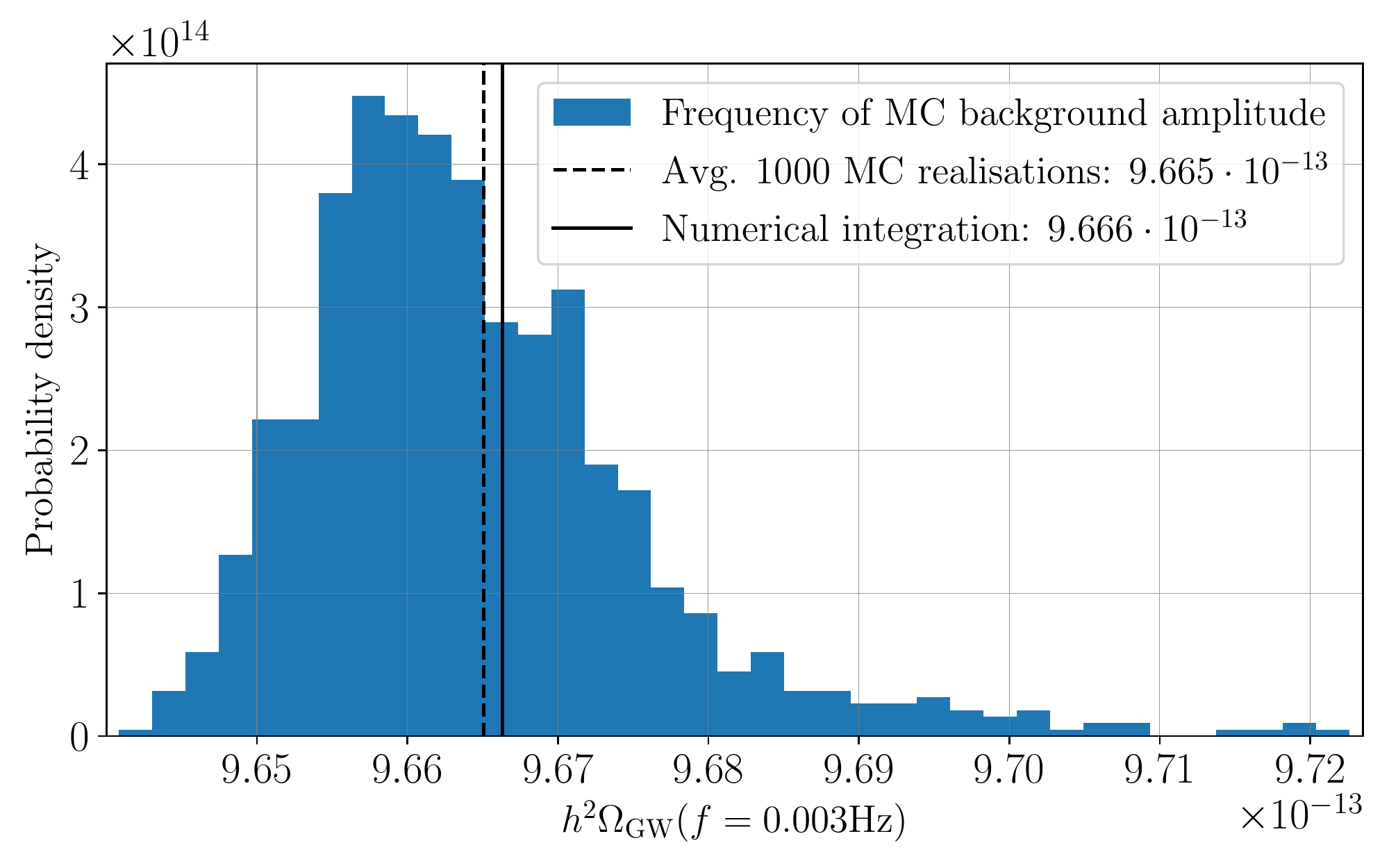}
  \caption{
    Comparison between the distribution of MC sums of \Cref{eq:char_strain_mcsum} over $1000$ realisations sharing the same population parameters (see \Cref{sec:fixedpoint}), and the numerical integration of \Cref{eq:CoeffOmegaGWAnalytic}. The two methods are equivalent within the range of the population variance ($\sim 0.2\%$), which is in turn much smaller than the integration error due to the $z_\mathrm{max}=5$ cut ($\sim 1\%$, see \Cref{fig:analyticalSignalII}). Such small effects will not be observable by LISA (see \Cref{sec:backgroundparamsestimation}).
    }
    \label{fig:char_strain_realisations}
\end{figure}

The SGWB amplitude resulting from the sum over a realisation of the SOBBH population is obviously realisation-dependent. 
We can assess its concordance, within the variance due to the population draws, with the analytical computation of \Cref{eq:CoeffOmegaGWAnalytic} by evaluating \Cref{eq:char_strain_mcsum} for a large number of realisations. 
The result is shown in \Cref{fig:char_strain_realisations}, assuming population parameters fixed to the benchmark values described in \Cref{sec:fixedpoint}, and setting $z_\mathrm{max}=5$. 
The population variance in terms of the ratio of the interquartile range to the mean of the realisations' amplitudes amounts to $0.2\%$ only.\footnote{If one naively computes the realisation variance as that of the underlying Poisson point process (i.e.\ equal to the mean of the process, here the expected number of events), one would overestimate the realisation uncertainty by a very large factor since not all events contribute equally to the SGWB.} 
The difference between the SGWB amplitude obtained by averaging the realisations, and the one obtained by the numerical integration of \Cref{eq:CoeffOmegaGWAnalytic}, is much smaller than the population variance, highlighting the equivalence between the two methods. 
Furthermore, the population variance uncertainty is much smaller than both the expected integration error due to fixing $z_\mathrm{max}=5$, and the forecasted precision of the  LISA measurement (see~\Cref{sec:backgroundparamsestimation}).

A more refined approach to evaluating the SGWB can be obtained by summing the contribution of each SOBBH in the population, accounting for the actual frequency of emission of each source (while in \Cref{eq:char_strain_mcsum} only the chirp mass and the distance - equivalently redshift - pertain to the individual events). 
In order to do this, we rewrite the SGWB energy density starting from \Cref{eq:omega},
but re-expressing the number density as the number of events per unit of emission frequency $f_s$ using the relation $\diff f_s/\diff \tau_c$ for quasi-circular Newtonian binaries, then changing the integration variable to the observed frequency, and equating the integrands in \Cref{eq:omega} to single out the SGWB power spectrum:
\begin{align}
  \Omega_{\rm GW}(f) = \int \diff \xi \int \diff z \, \frac{\diff^3N(z, \xi,\theta)}{\diff\xi \diff z \,\diff f} \,f \frac{1}{4\pi} \int \diff\Omega\, \frac{\rho_\mathrm{GW}^{\rm (event)}}{\rho_c}\,.
\end{align}
We can now express the integral in the above equation as an MC sum, as done previously in \Cref{eq:char_strain_mcsum}, but this time computing the sum of the GW energy density emitted by every SOBBH per (detector-frame) frequency bin, where the latter is defined using some frequency sampling $\delta_f$ as $[(j-1)\delta_f, j\delta_f]$, $N_j$ being the subset of a population with emission frequencies (in detector frame) in bin $j$
\begin{equation}
\begin{split}
   \Omega_{\rm GW}(f) &\approx \frac{1}{\delta_f} \sum_{i\in N_j}
   f_{i} \frac{1}{4\pi} \int \diff\Omega\,\frac{\rho_\mathrm{GW}^{\rm (event)}(z_i, \mathcal{M}_i, f_{i})}{\rho_c}\\
   &\approx \frac{64\pi^{10/3} }{15} \frac{G^{10/3}}{c^8H_0^2} \frac{1}{\delta_f} \sum_{i\in N_j}
   \frac{(1+z_i)^{4/3}}{d_{M, i}^2} \mathcal{M}_i^{10/3} f_i^{13/3}
   \,,
   \label{eq:omega_mc_2}
   \end{split}
\end{equation}
where the different powers of the per-source quantities with respect to \Cref{eq:char_strain_mcsum} can be explained by the frequency dependence of the number of sources in each bin.

This assumes monochromatic sources, ignoring frequency drifting during the life of the mission.\footnote{We could easily extend this formula to account for drifting by summing each event over different bins with some weight proportional to the time spent emitting at the frequency of the bin \cite{2020PhRvD.102j3023B}.
} 
The largest contribution to the background is produced by sources with $f\in(10^{-3},10^{-2})$, whose frequency drifting is small; we can therefore choose e.g.\ the frequency with which they enter the LISA band (see \Cref{app:ttot}). 
We will show the result of both MC integrations, Eqs.~\eqref{eq:char_strain_mcsum} and \eqref{eq:omega_mc_2}, in \Cref{sec:comparison}.

\subsection{Method (iii): iterative subtraction}
\label{sec:iterative}

The methods presented above are based on summing the signals of the SOBBH in the population, without accounting for the actual detection process, apart from restricting the maximal time-to-coalescence $\tau_{\rm c, max}^{\rm (det)}$ to a computationally manageable and detector-compatible value (for methods (iia) and (iib) of \Cref{sec:montecarlo}). 
However, we are ultimately interested in the SGWB signal in LISA, and the detector sensitivity can influence the SGWB spectral shape/amplitude.  
In order to consider such aspects, we also evaluate the SGWB following the methodology developed in \cite{Karnesis:2021tsh}, using ideas first presented in~\cite{Timpano2006, crowder2007,  Nissanke2012eh}. 
The procedure is based on generating LISA data-streams, by computing the waveform signals of all the events within the simulated population. 
Depending on the adopted waveform model, this can yield a very accurate representation of the LISA data, as far as SOBBHs are concerned. 
However, simulating millions of sources is computationally expensive, thus one has to allocate a considerable amount of computational resources to this task.

The procedure begins by fixing the mission duration $T_\mathrm{obs}$, here set to 4 years, and generating the signal to be measured by LISA. 
We compute the $h_+$ and $h_\times$ waveforms for each source of the simulated catalogue, and then we project them onto the LISA arms. 
We use the {\tt IMRPhenomHM} model~\cite{London:2017bcn}, which describes spinning, non-precessing 
binaries. 
It is based on the {\tt IMRPhenomD}~\cite{2016PhRvD..93d4006H, 2016PhRvD..93d4007K} model, but it includes higher order modes. 
We use the {\tt lisabeta} software~\cite{lisabeta_soft, Marsat:2018oam} for our computations. 
When generating each waveform, we also compute their 
SNR in isolation, $\rho^\mathrm{iso}_i$, with respect to the instrumental noise only, which will be used to reduce the computational requirements of the procedure, as explained below.

Next, we estimate the total power spectral density (PSD), $S_{ \rm n,\,k}$, summing all the GW sources plus the instrumental noise. 
The index k refers to the iterative step. 
Since this PSD is very noisy, we compute its running median to produce a smoother version of it. 
We 
then evaluate the SNR $\rho_i$ of each source $i$ using the smoothed $S_\mathrm{n,\,k}$ as the total ``noise'' PSD. 
Note that, to speed up the computation, this is performed only on the subset of sources with sizable pre-computed SNR in isolation $\rho^\mathrm{iso}_i$ (see \cite{Karnesis:2021tsh}). 
The SNR $\rho_i$ are then compared to a threshold SNR $\rho_0$: if $\rho_i > \rho_0$, the source is classified as resolvable, and is subtracted from the data.
The smoothed residual PSD $S_\mathrm{n,\,k+1}$ is then re-evaluated after re-iterating through the catalogue of sources and subtracting the loud ones, 
and the procedure is repeated until the algorithm converges. 
Convergence is reached when all the sources are subtracted given the $\rho_0$ threshold, or if $S_\mathrm{n,\,k+1}$ and $S_\mathrm{n,\,k}$ are practically identical at all frequencies considered. 
At the end of the procedure, we compute the final SNR of the recovered sources, with respect to the final estimate of $S_{ \rm n,\,k_\mathrm{final}}$.
Thus, as final products, we get both the SGWB due to the sources signal confusion, as well as the properties of the recovered sources (their number, waveform parameters, and final SNR). 

\begin{figure}[t!]
    \centering
    \subcaptionbox{%
    \label{fig:statseffect}}%
    {\includegraphics[width=.46\textwidth]{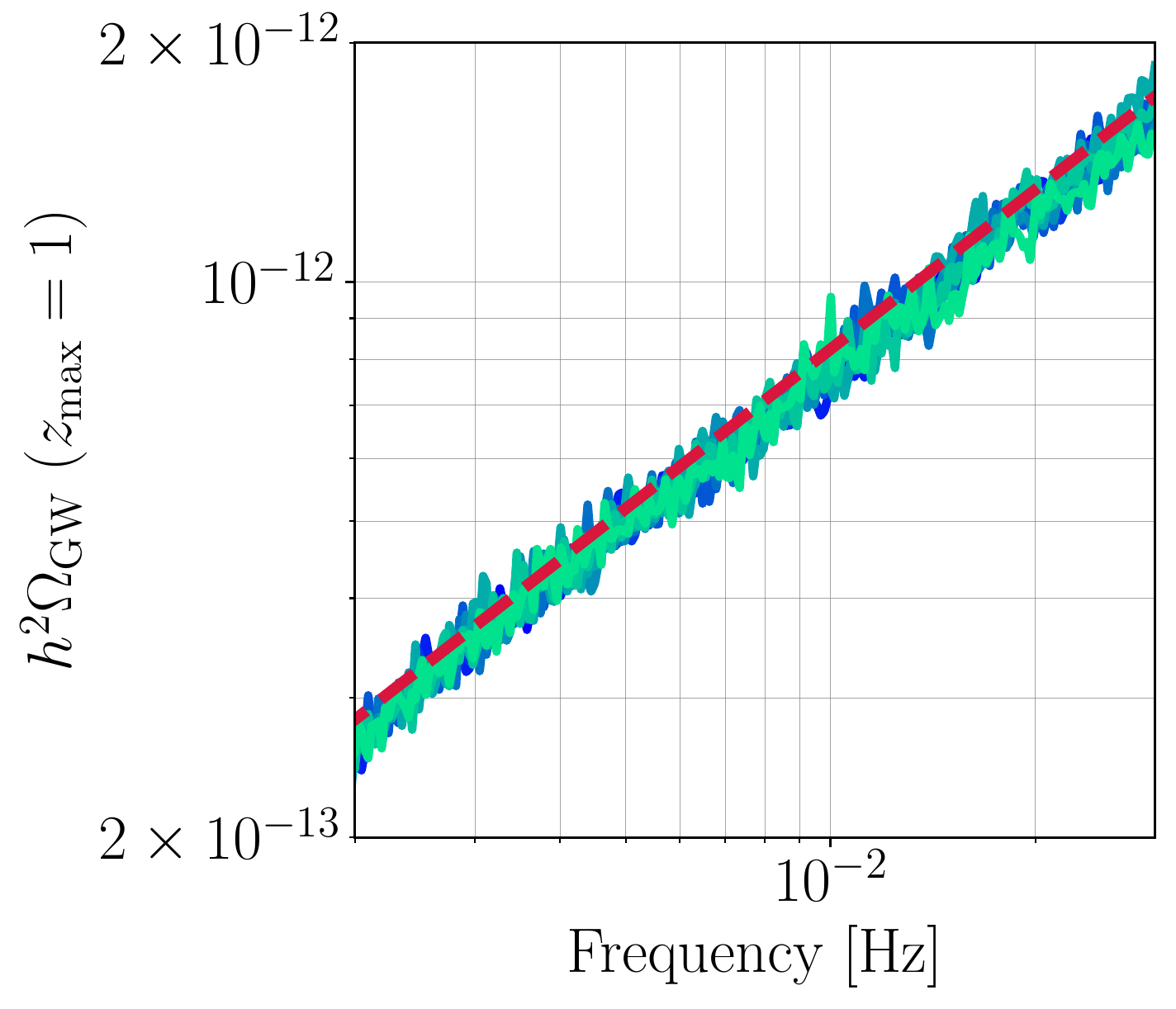}}
    \subcaptionbox{%
    \label{fig:snrseffect}}%
    {\includegraphics[width=.425\textwidth]{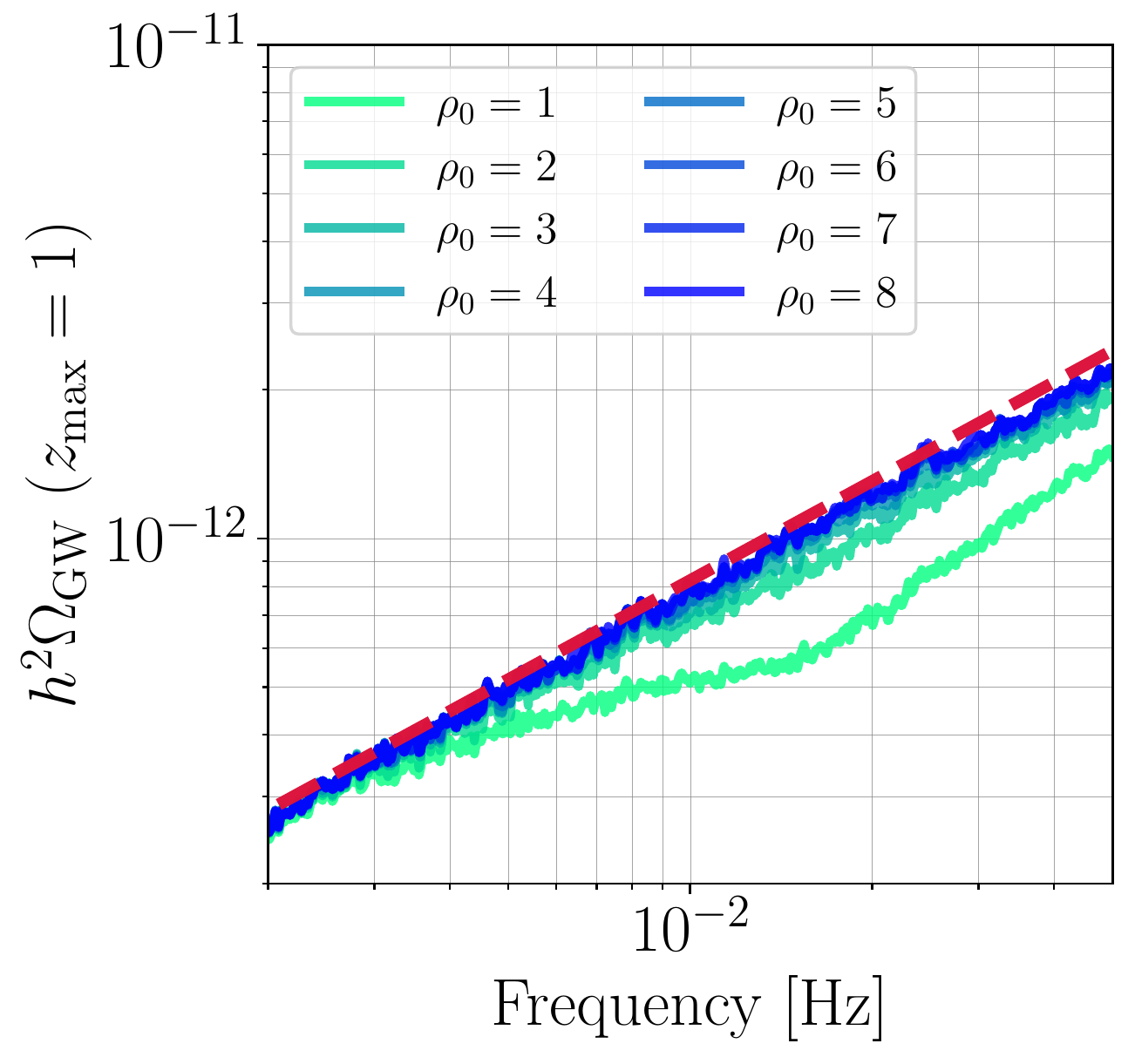}}%
  \caption{Tests of the iterative-subtraction method: the SOBBH SGWB is shown in comparison with the analytical evaluation obtained integrating  \Cref{eq:CoeffOmegaGWAnalytic} (red dashed line) for the benchmark population parameters described in \Cref{sec:fixedpoint}, simulated, and integrated, up to $z_\mathrm{max}=1$: (\subref{fig:statseffect}) the 10 realisations of the simulated population with equal parameters, set to the benchmark described in \Cref{sec:fixedpoint}, yield the same level of stochastic signal; (\subref{fig:snrseffect}) the effect of different choices for the SNR threshold $\rho_0$ on the resulting stochastic signal: the smaller $\rho_0$, the more sources are classified as resolvable, generating a ``dip'' in the stochastic signal at high frequencies. The shape of this dip depends on both the astrophysical catalogue, and on the shape of the instrumental noise PSD.
  }
    \label{fig:iterativechecks}
\end{figure}

Different realisations of the same population (with the same number density parameters) should yield different, though statistically compatible, sets of subtracted events, but a similar SGWB after smoothing. 
We have verified this statement by evaluating the SGWB from the 10 benchmark catalogues presented in \Cref{sec:fixedpoint}; the result is shown in \Cref{fig:statseffect}.

The crucial parameter of the iterative method is $\rho_0$, the minimum SNR above which events are considered resolvable and thus subtracted from the total signal. 
We consider $\rho_0=8$ an appropriate choice \cite{Buscicchio:2021dph,Toubiana:2022vpp}, assuming that stochastic methods to sample the sources parameter space, more efficient than grid-based methods \cite{Moore:2019pke}, can be used to analyse the LISA data streams.
Archival searches will allow to further reduce the SNR threshold down to $\rho_0= 5$ \cite{Wong:2018uwb,Toubiana:2022vpp}.
As can be appreciated from \Cref{fig:snrseffect}, as long as $\rho_0\gtrsim 5$, the number of detectable sources is too small to alter the shape and amplitude of the residual SGWB spectrum \cite{Sesana:2016ljz,Perigois:2020ymr} (see also \cite{next}).
Our prediction for the SGWB level is therefore robust with respect to our choice of setting $\rho_0=8$.
On the other hand, if values of $\rho_0\lesssim 4$ will be justified in the context of future improvements in data analysis methods, or of archival searches using future ground-based detector data \cite{Regimbau:2016ike}, the residual SGWB spectral shape must be adapted: as can be seen in \Cref{fig:snrseffect}, it no longer follows the analytical estimation of \Cref{sec:analytic}, which does not account the presence of the detector, but
a dip on its amplitude appears at high frequencies.

Note that we have assumed uninterrupted measurement over the time frame $T_\mathrm{obs}$, and the instrumental noise, taken from \cite{SciRD, Babak2021mhe}, is assumed to be ideal, i.e.~Gaussian and stationary. 
We also subtract each resolvable source from the data at its injection parameters, meaning that we generate ``perfect residuals'', or in other words, we neglect 
the uncertainty on the source parameters, which inevitably arises within the parameter estimation procedure. 
We, therefore, simulate an optimal case of the {global fit} scheme for the {LISA} SOBBHs. 
The above assumptions, while not totally realistic, allow us to simplify the analysis.

\section{Results}
\label{sec:results}

\subsection{Comparison between SGWB computation methods in the LISA band}
\label{sec:comparison}

In this section we 
show the effect of fixing a maximal time-to-coalescence for the simulated populations on the SGWB spectral shape, and compare the SGWB signals resulting from the four methods described in Sections \ref{sec:analytic} to \ref{sec:iterative}. 
As a benchmark, we use one of the fixed-point catalogues presented in  \Cref{sec:fixedpoint}. 
The redshift range is limited to $z \in [0, 1]$ (comoving distance up to $\approx 3~\rm{GPc}$) to guarantee the computational feasibility of the iterative-subtraction method. 
The amplitude of the SGWB signals shown in this section is therefore reduced (cf.~\Cref{sec:postpredictive}), 
but this plays no role in the purpose of the tests performed here.

\begin{figure}[!htb]
    \centering
    \includegraphics[width=.45\textwidth]{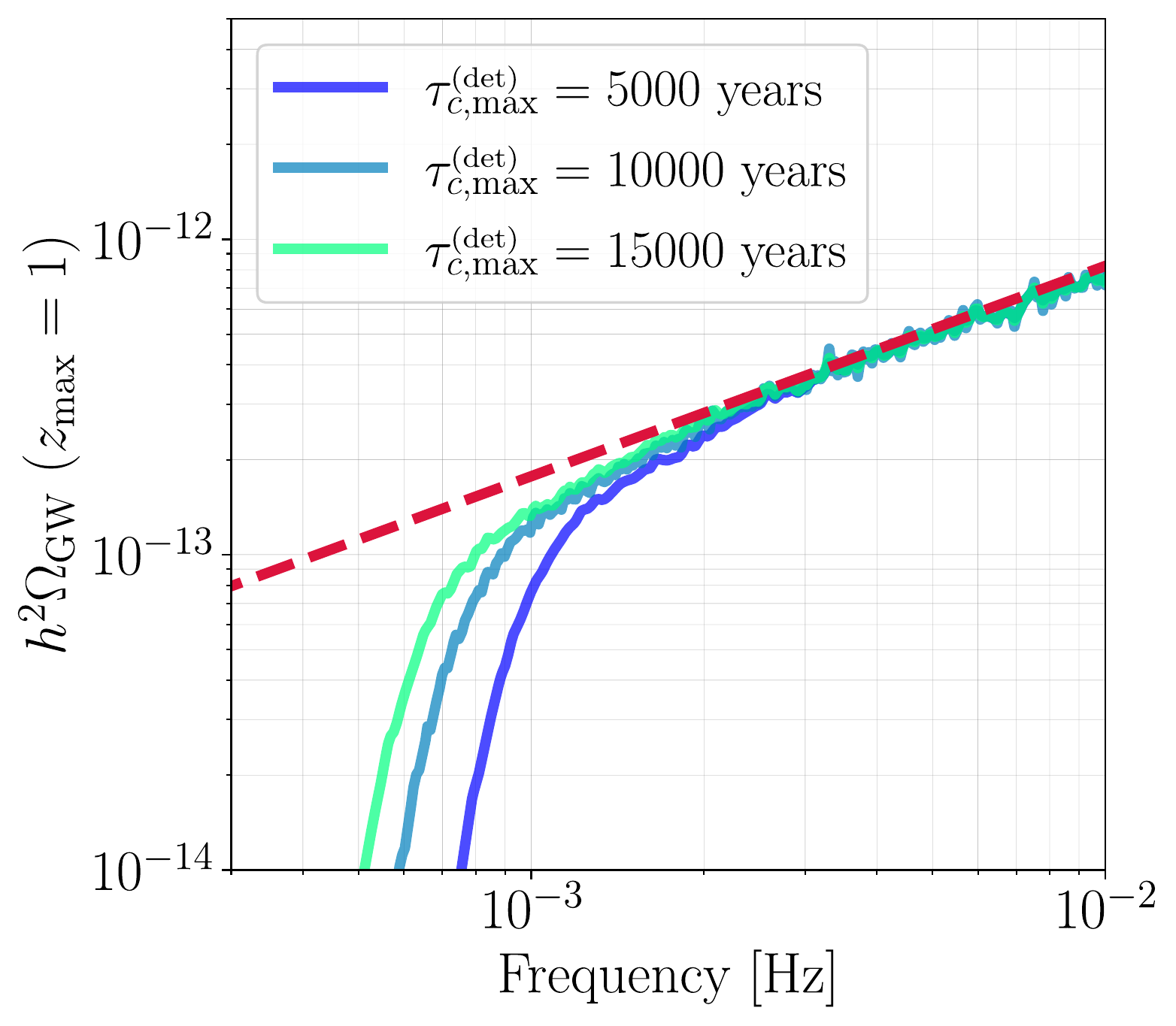}
  \caption{The effect of adopting different $\tau_{c,\mathrm{max}}^{(\mathrm{det})}$ values on the resulting stochastic signal, computed using the iterative subtraction (method (iii)). 
  The red dashed line represents the analytical result (method (i)).
  Imposing a maximum time-to-coalescence in generating the synthetic populations suppresses early-phase inspirals, producing a cutoff in the SGWB at low frequencies. 
  This is not a physical effect, but a limitation of the population synthesis: the spectrum tends towards the expected power law as the upper limit in time-to-coalescence grows.}
    \label{fig:tceffect}
\end{figure}

As discussed in Sections \ref{sec:popsynthesis} and \ref{sec:fixedpoint}, in order to limit computational costs, synthetic populations are generated including events up to a maximum time-to-coalescence, that we fix to $\tau_{c,\mathrm{max}}^{(\mathrm{det})}=10^4\,\yr$ in the detector frame. 
In order to investigate the effect of this assumption, one of the catalogues among the benchmark ones has been generated with $\tau_{c, \mathrm{max}}^{(\mathrm{det})}=1.5\times10^4\,\yr$, and from it we have produced two sub-catalogues with $\tau_{c, \mathrm{max}}^{(\mathrm{det})}=1.0\times10^4$ and $5\times10^3\,\yr$. 
The SGWBs inferred from these catalogues via the iterative-subtraction method are shown in \Cref{fig:tceffect}:
excluding all sources beyond a given $\tau_{c, \mathrm{max}}^{(\mathrm{det})}$ (appropriately redshifted in the source frame), results in a non-physical bending of the SGWB at low frequencies, depending on the maximal time-to-coalescence (in agreement with~\cite{Karnesis:2021tsh}, see also \Cref{app:ttot}). 
It is therefore important to pick a value for $\tau_{c, \mathrm{max}}^{(\mathrm{det})}$ ensuring a minimal loss of information while keeping the computational cost of generating the SGWB manageable: as discussed in \Cref{app:ttot}, we consider $\tau_{c\mathrm{max}}^{(\mathrm{det})}=10^4$ to be a good compromise.

\begin{figure}[!htb]
    \centering
    \includegraphics[width=.72\textwidth]{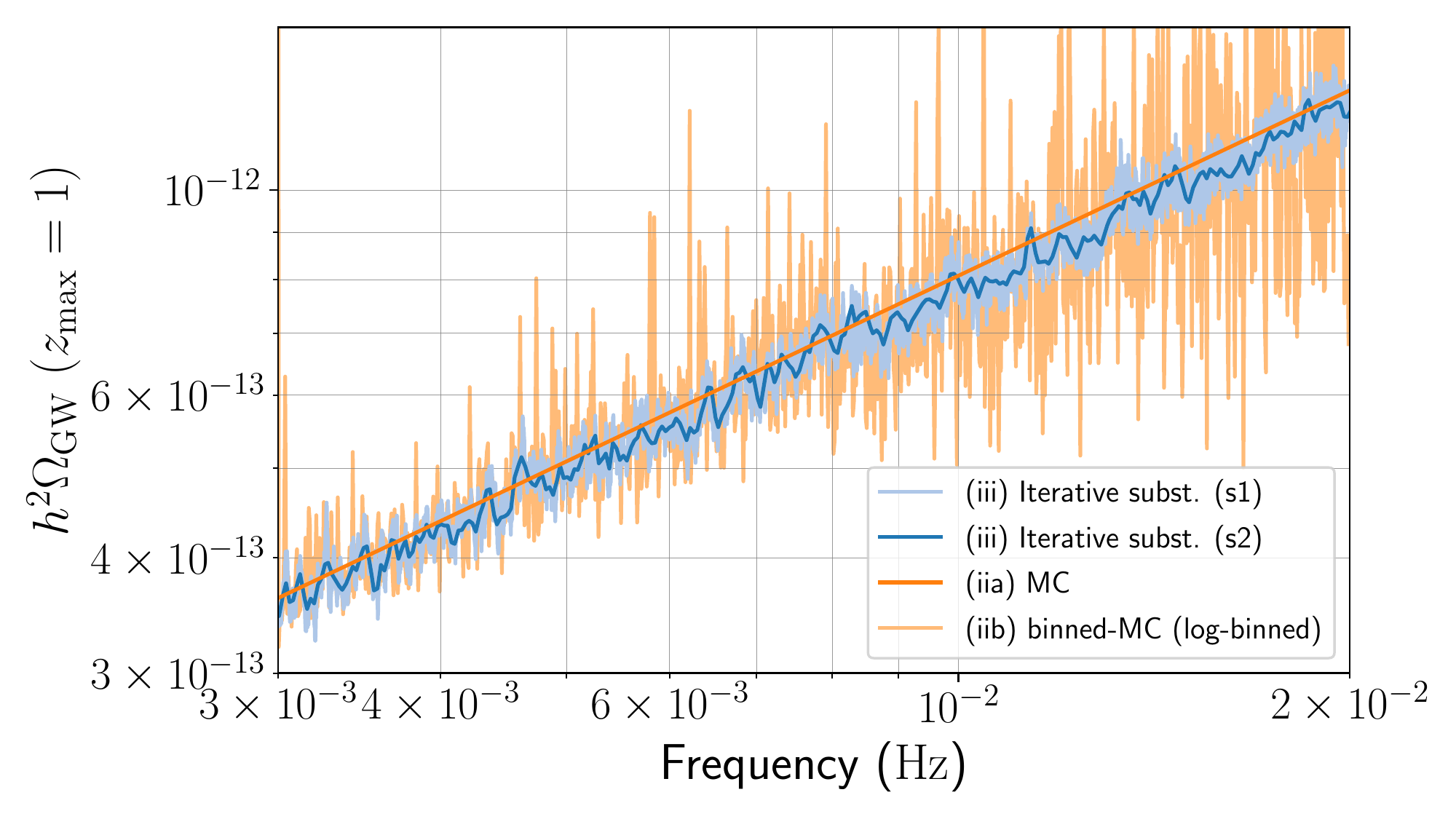}
  \caption{
  Comparison between the population synthesis-based methods presented in Sections \ref{sec:montecarlo} and \ref{sec:iterative}. The dark orange line shows the SGWB evaluated with the averaged MC sum (iia), and the light orange one the frequency-binned sum (iib). 
  The blue curves show the SGWB evaluated with the iterative subtraction (iii),~\Cref{sec:iterative}, for two different data smoothing methods: in s1 (light blue) we have performed a running median over the PSD data using a rolling window of 1000 points, whereas in s2 (dark blue) we apply an additional Gaussian filter. The signals from the frequency-binned MC sum and from the iterative subtraction share some features, especially at low frequencies, where the differences due to neglecting the drifting and using simplified waveforms are less important. Both follow closely the SGWB of the averaged, power-law-like MC sum.}
    \label{fig:comparison}
\end{figure}

In \Cref{fig:comparison} we show the SGWBs computed using the three methods based on population synthesis, presented in Sections \ref{sec:montecarlo} and \ref{sec:iterative}. 
The results are in very good agreement, for both the SGWB amplitude and spectral shape. 
In particular, those of the frequency-binned MC sum (method (iib)) and of the iterative subtraction (method (iii)) also follow the single power-law behaviour $f^{2/3}$ predicted by the analytical evaluation (\Cref{eq:Omega_int}), and taken over by the averaged power-law-like MC sum (\Cref{eq:char_strain_mcsum}).  
As far as the frequency-binned MC sum is concerned, this shows that our population catalogues are complete. 
As far as the iterative-subtraction method is concerned, instead, this is a consequence of the simulated detection process: the instrument sensitivity is such that the number of resolvable sources is too small, even at high frequencies, to alter the SGWB spectral shape, as already pointed out in \cite{Perigois:2020ymr} (see also \cite{next}). 

The signals from the frequency-binned MC sum and from the iterative subtraction share some features, especially at low frequencies, despite the fact that the former uses simplified waveform and does not account for frequency drifts. 
Both approaches also follow closely the averaged power-law-like MC sum, which is distributed around the analytical calculation of the background, from \Cref{eq:CoeffOmegaGWAnalytic} (see \Cref{fig:char_strain_realisations}).

\subsection{Expected SOBBH signal in the LISA band from GWTC-3}
\label{sec:postpredictive}

\begin{figure}[t!]
    \centering
    \includegraphics[width=.9\textwidth]{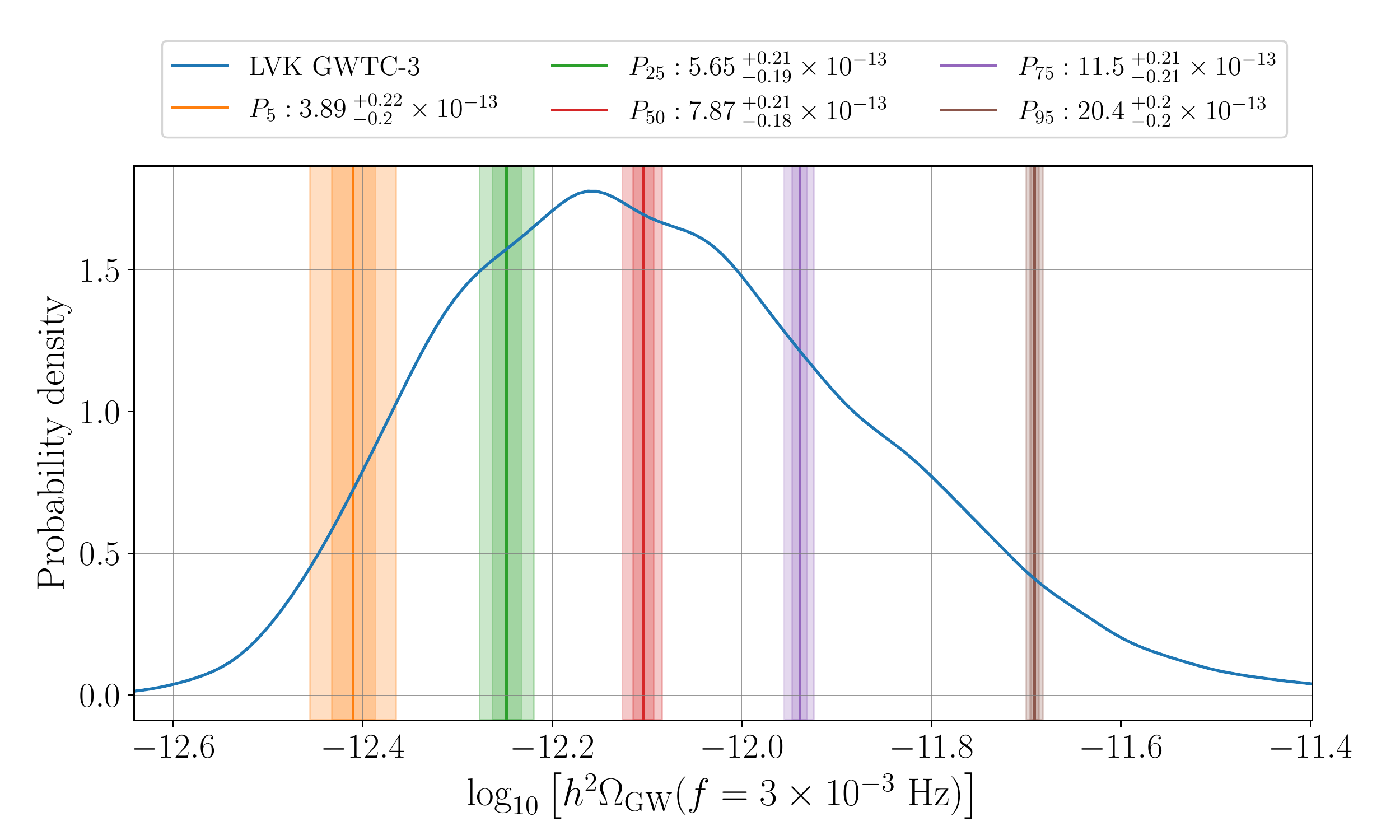}
  \caption{
    Posterior distribution of the SOBBH SGWB log-amplitude in the LISA band, from GWTC-3 (blue solid line). 
    The coloured lines represent the percentiles 5, 25, 50, 75, and 95 (left to right) of the posterior, and their surrounding  
    vertical bands represent the $68\%$ (dark shading) and $95\%$ (light shading) uncertainties on the corresponding SGWB amplitudes, forecasted from a LISA measurement (the uncertainties quoted in the legend correspond to the $68\%$ error): as derived in \Cref{sec:backgroundparamsestimation}, LISA will measure the SOBBH SGWB with an uncertainty on the amplitude one order of magnitude smaller than the present GWTC-3 prediction.}
    \label{fig:bkgperc}
\end{figure}

Having established the consistency of the four methods, we turn to the actual computation of the expected SGWB in the LISA band, based on the present knowledge about the SOBBH population. 
To this purpose, we rely on \Cref{eq:powerlaw} and evaluate the SGWB amplitude by integrating \Cref{eq:CoeffOmegaGWAnalytic} for all points in the LVK posterior parameter sample that is publicly available \cite{ligo_scientific_collaboration_and_virgo_2021_5655785} for the \FID model \cite{LIGOScientific:2021psn}, following the prescriptions described in \Cref{sec:postparams}. 
The distribution of the SGWB amplitude at the reference frequency $f=3\times10^{-3}\,\mathrm{Hz}$ is shown in \Cref{fig:bkgperc} (blue solid line). 
On a logarithmic scale, it follows a lightly-right-skewed distribution with median $\omhsqpivot = 7.87\times10^{-13}$, and has an interquartile range of $\omhsqpivot \in [5.65,\,11.5]\times10^{-13}$.

\begin{figure}[t!]
    \centering
    \includegraphics[width=.8\textwidth]{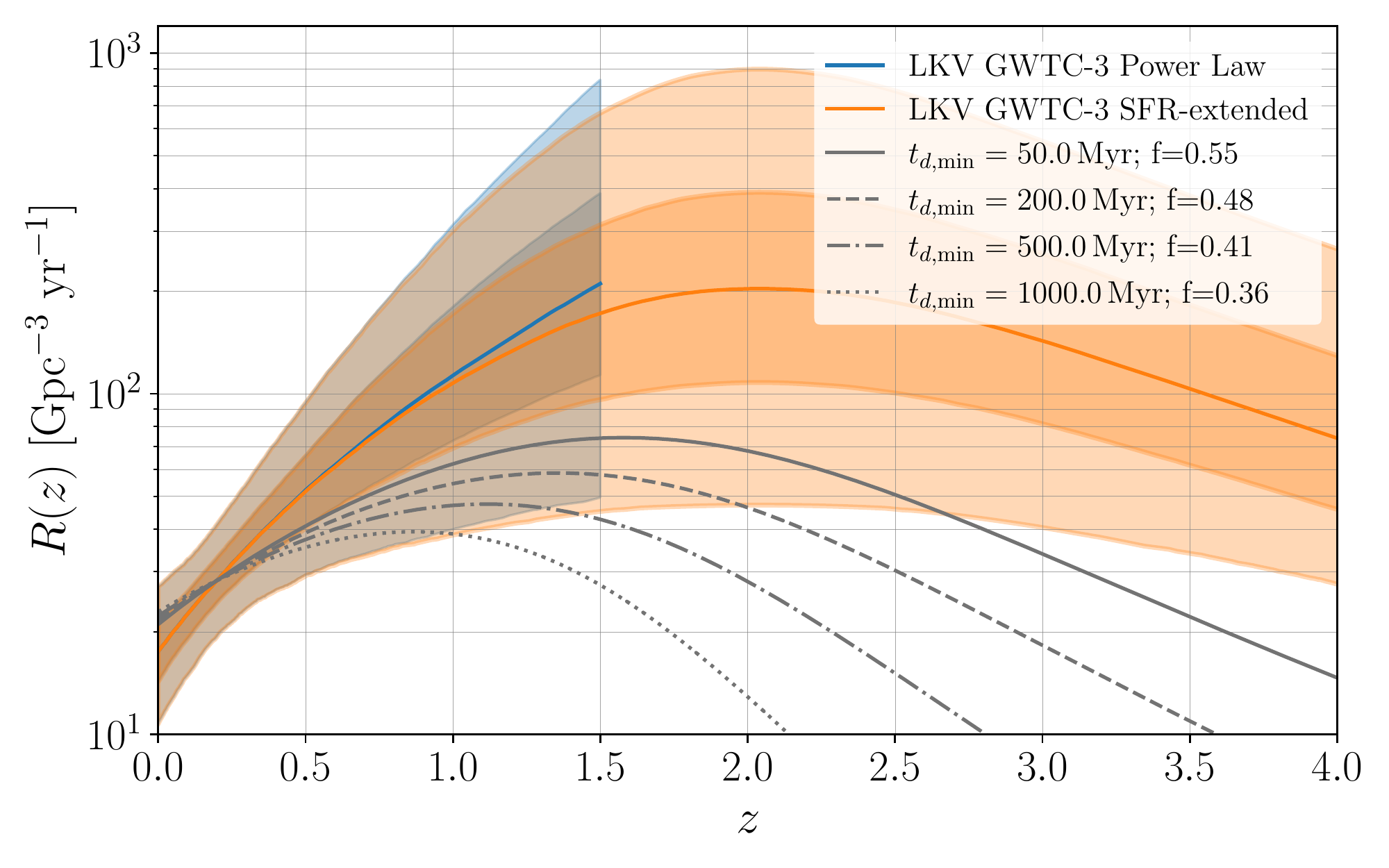}
  \caption{Blue: probability distribution (median, $25$--$75$ and $5$--$95$ percentile ranges) of the merger rate under the low-z power-law approximation $R(z) \propto (1+z)^\kappa$, according to the probability density obtained in \cite{LIGOScientific:2020kqk} for the \FID model. Orange: its extension to high-$z$ according to \Cref{eq:rz}, with fiducial values $z_\mathrm{peak}=2.04$, $r=3.6$.
  Gray lines: the SOBBH merger rate obtained by convolving the SFR with a time delay, for different values of the minimum time delay, \Cref{eq:merger_td}. 
  For each case in the legend, f represents the fraction of the corresponding SGWB amplitude with respect to the median fiducial case (obtained using  \Cref{eq:rz} rather than \Cref{eq:merger_td}). 
  }
    \label{fig:merger_rate_delay}
  \end{figure}

The computation of the SGWB amplitude has been performed under the assumption that the merger rate inherits the functional redshift dependence of the SFR, \Cref{eq:rz}.
As discussed in \Cref{sec:popmodel} and \Cref{app:mergerrate}, the agreement between the values of the merger rate parameters inferred from GW observations \cite{LIGOScientific:2021psn} with those of the SFR inferred from electromagnetic observations \cite{Madau:2016jbv} supports this assumption at low redshift $z\lesssim 1.5$. 
However, the merger rate remains untested at higher redshifts, and it is therefore important to investigate how much this assumption influences the final SGWB result.  
We do so by analysing one example of a more refined model for the merger rate, introducing a time delay $t_d$ between the formation of star binaries and their evolution into BBH systems. 
The merger rate is then given by the convolution of the SFR with the probability distribution of the time delay \cite{Dvorkin:2016wac,Santoliquido_2020,Fishbach_2021,van_Son_2022}:
\begin{equation}
R(z) = \int_{t_{d,\mathrm{min}}}^{t_{d,\mathrm{max}}} R_\mathrm{SFR}\left(t(z) + t_d\right) p(t_d)\,\mathrm{d}t_d\,,
\label{eq:merger_td}
\end{equation}
where $R_\mathrm{SFR}$ now denotes the rate of \Cref{eq:rz} (with parameter values specified below that equation), and the probability distribution of the time delay is usually modelled as $p(t_d) \propto 1/t_d$. 
As a consequence, the minimum expected delay $t_{d,\mathrm{min}}$ plays the main role in determining the merger rate dependence on redshift.
This parameter is expected to lay in the range $50\,\mathrm{Myr}$--$1\,\mathrm{Gyr}$ \cite{Dvorkin:2016wac}: we therefore pick four values in this range, and compute the  corresponding merger rates from \Cref{eq:merger_td}, further imposing that at $z=0.2$ they are equal to the median value of the GWTC-3 constrain for the \FID model, $R_{0.2}={28.3}\,\Runits$ (see \Cref{fig:merger_rate_delay}) \cite{LIGOScientific:2021psn}. 
Looking at \Cref{eq:CoeffOmegaGWAnalytic}, we see that (for redshift-independent mass models) the redshift-dependent contribution to the background amplitude can be factored out. We can thus easily compute, for a given mass model, the ratio $f$ between a time delayed model and our fiducial SRF case.
For $t_{d,\mathrm{min}}=50$ to $500\,{\rm Myr}$, we find that they agree within $40\%$:  accounting for the time delay, therefore, provides SGWB amplitudes close to the $P_{5}$ percentile of the median fiducial (SFR-extended) case (see \Cref{fig:bkgperc}).
The level of agreement drops to 36\% for $t_{d,\mathrm{min}}=1$ Gyr;
however, from \Cref{fig:merger_rate_delay}, we can appreciate that the corresponding merger rate is rather in tension with LVK constraints. 

Our results, in terms of translating the GWTC-3 population constraints into a forecast for the SOBBH SGWB in the LISA band, appear to be robust within one order of magnitude:
the highest contribution to the background comes in fact from  the SOBBH population at $z\lesssim 1.5$, for which the merger rate is well constrained by LVK GWTC-3. 
Note that all derived SGWB amplitudes fall well within LISA's detection capabilities (see \Cref{sec:backgroundparamsestimation}). 
A more thorough study of the dependency of the SGWB amplitude on physically-motivated models for the merger rate and mass distribution can be found in \cite{newIrina}.

\begin{figure}[t!]
    \centering
    \includegraphics[width=.9\textwidth]{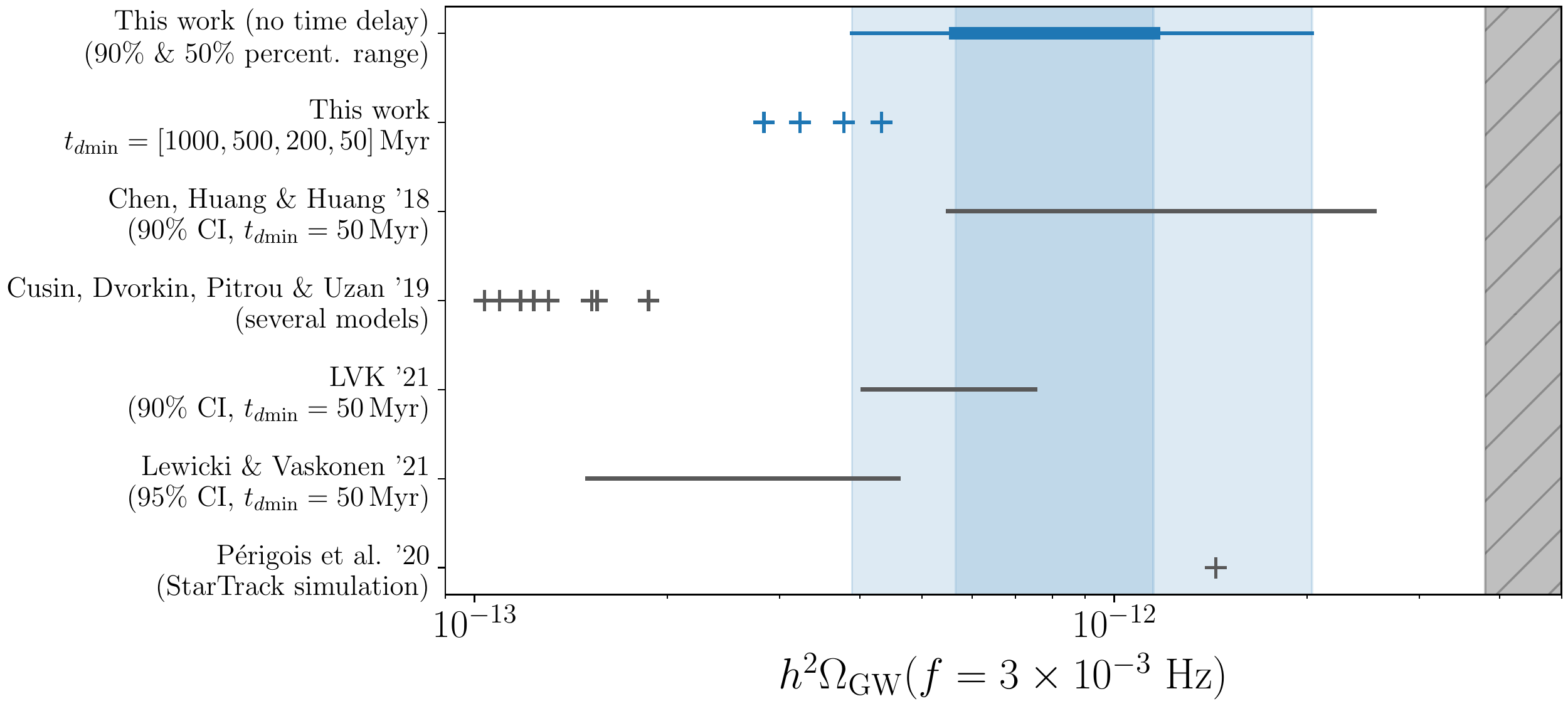}
  \caption{Comparison between the SGWB amplitude posterior from this work (blue shaded area, also shown in \Cref{fig:bkgperc}), with the median SGWB value evaluated accounting for time-delays (blue crosses, left-to-right in decreasing value of $t_{d,\mathrm{min}}$, cf.~\Cref{fig:merger_rate_delay}), and with other recent results from the literature (black lines and crosses). The grey band to the right represents the LVK upper bound, derived in \cite{KAGRA:2021kbb}  
  for a power-law SGWB with index $\alpha=2/3$.
  }
    \label{fig:background_comparison}
  \end{figure}

Our results are also compatible with the latest constraints on the SGWB amplitude by LVK \cite{KAGRA:2021kbb}.
The upper bound on a power-law SGWB with spectral index $2/3$ at $f=25\,\mathrm{Hz}$ is,  at $95\%$ credible level, $3.4\times10^{-9} (1.2\times10^{-8}$), when using a log-uniform (uniform) prior, which becomes in the LISA band $\omhsqpivot<3.8\times10^{-12}\,(1.3\times10^{-11})$. 
This upper bound applies to the total background, which contains other contributions together with the SOBBH confusion signal (for example the one from neutron star binaries). 
The actual limit on the SOBBH SGWB is therefore expected to be smaller.
Nevertheless, the SGWB amplitude that we forecast remains compatible, being smaller than the LVK upper limit at 99\% probability, with median amplitude being smaller by a factor of five (see \Cref{fig:bkgperc}).

We also compare our results to a few other predictions for the SOBBH SGWB in the LISA band given in the literature, see \Cref{fig:background_comparison}.
In \cite{Chen:2018rzo}, taking into account early LVK constraints (from the first 6 events) for the merger rate, a time delay distribution $p(t_d) \propto 1/t_d$ with $t_{d,\mathrm{min}}=50\,\mathrm{Myr}$, and a different fiducial model for the mass distribution from the one used here, it was found that $\omhsqpivot=\pmint{1.25}{0.7}{1.3}\times10^{-12}$ ($90\%$ credible level), which lies in the upper-half of our distribution (see \Cref{fig:bkgperc}). In \cite{Cusin:2019jpv}, the authors compute both the isotropic SOBBH SGWB component and its anisotropy, and find a lower prediction than in our analysis: 
$\omhsqpivot\in[1.0, 1.9]\times10^{-13}$,
for a number of astrophysics-motivated models for the merger rate, adjusted to LVK GWTC-1 constraints.
The latest LVK forecast \cite{KAGRA:2021kbb}, using the merger rate and the mass distribution inferred from GWTC-2, and the usual time-delay distribution, results in $\omhsqpivot=\pmint{5.6}{1.6}{1.9}\times10^{-13}$ ($90\%$ credible level), which is consistent with our results both when including and not including time delays. 
The analysis of~\cite{Lewicki:2021kmu}, also based on the LVK GWTC-2 population model, uses power-law mass functions and the conventional time-delay distribution, and obtains $\omhsqpivot=\pmint{2.9}{1.4}{1.7}\times10^{-13}$ (approx $95\%$ credible level): this prediction  is compatible with our results, but towards the low side of the distribution in \Cref{fig:bkgperc}.
In~\cite{Perigois:2020ymr}, the authors use the population code Star-Track to model the binary formation, treating separately the contributions from population I/II and population III stars.
The SGWB amplitude from SOBBHs formed by population I/II star is $\omhsqpivot=1.2\times10^{-12}$, which lays in the upper part of our probability distribution. 
Population III stars contribute an additional 2\%, $\omhsqpivot=2.25\times10^{-13}$:
since this is significantly larger than the expected uncertainty in LISA's measurement of the background (see \Cref{sec:backgroundparamsestimation}), the presence of population III stars could be discriminated, provided that the population is known with sufficient certainty.
Finally, in \cite{Bavera:2021wmw} it is found that the contribution of SOBBH to the SGWB is even lower than what found in \cite{Cusin:2019jpv}, and subdominant in the LISA band with respect to the one from primordial BHs: $\omhsqpivot \simeq 4.5\times 10^{-14}$.

\subsection{SGWB Parameter Estimation}
\label{sec:backgroundparamsestimation}

In this section we assess
LISA's capability to detect and characterise the SOBBH SGWB.
We perform an MC analysis 
of simulated data containing the instrumental noise, the stochastic foreground from binaries in the Galaxy, and different levels of the SOBBH SGWB,
corresponding to the percentiles presented in \Cref{fig:bkgperc}.
The SOBBH SGWB is modelled following 
\Cref{eq:powerlaw}, 
but both the amplitude and the spectral tilt are left as free parameters in the analysis:
\begin{equation}
    h^2 \Omega_\mathrm{GW}(f) = 10^{\log_{10}(h^2\Omega_\mathrm{GW})(f_*)}\left(\frac{f}{f_{*}}\right)^{\alpha}\,.
    \label{eq: SOBBH_for}
\end{equation}
We apply a pre-processing procedure similar to the one employed in~\cite{Caprini:2019pxz, Flauger:2020qyi}, which we briefly summarize here: assuming a mission duration of 4 years, we chunk the data stream into $N_c$ segments of $11.5$ days each (corresponding to a frequency resolution $\Delta f\simeq 10^{-6}\ \rm{Hz}$); we generate data in the frequency domain for each segment, including  
the instrumental noise, the GB foreground, and the SOBBH SGWB, and we average over these segments to get the simulated measured spectrum. 
Using the noise as an estimate for the variance, we define a likelihood consisting of a sum of Gaussian and log-normal components (the latter accounting for the skewness of the exact likelihood), as discussed in~\cite{Flauger:2020qyi}. 
For the sake of speed and without loss of precision, this likelihood is applied to a coarse-grained version of the spectrum obtained by inverse variance weighting, the final data in frequency space being defined as
\begin{equation}
    D_{ij}^{th}\left(f_{ij}^k\right) =  h^2 \Omega_\mathrm{GW}(f_{ij}^k, \vec{\theta}_s) +  h^2 \Omega_{n}(f_{ij}^k, \vec{\theta}_n) \,,
\end{equation}
where $f_{ij}^k$ and $ D_{ij}^k$ are the coarse-grained frequencies and data respectively. 
$\Omega_\mathrm{GW}$ represents both the SOBBH component, with spectral shape defined by \Cref{eq: SOBBH_for}, and the GB foreground component, based on the model from \cite{Karnesis:2021tsh}. 
$\vec{\theta}_{s}$ is the vector of parameters of the signal: amplitude and spectral tilt of the SOBBH SGWB, while we reconstruct only the amplitude $h^2\Omega_\mathrm{Gal}$ of the GB foreground.
$\Omega_{n}$ is the instrumental noise in omega units.  
We adopt a two-parameter noise model as typically done for LISA: the noise is characterized at low frequency by the acceleration component, parameterised by $A$, and at high frequency by the interferometric component, parameterised by $P$ \cite{Caprini:2019pxz}. 
The two noise parameters form the vector $\vec{\theta}_n$, and vary freely in our analysis.
We sample over the joint $(\vec{\theta}_{s}, \vec{\theta}_{n}) = (\log_{10}[\omhsqpivot], \alpha, \log_{10}[h^2\Omega_\mathrm{Gal}], A, P)$ parameter space using the Nested Sampler \texttt{Polychord}~\cite{Handley:2015fda,2015MNRAS.453.4384H} via its interface with \texttt{Cobaya}~\cite{Torrado:2020dgo}.

\begin{figure}[t!]
    \centering
    \includegraphics[width=\textwidth]{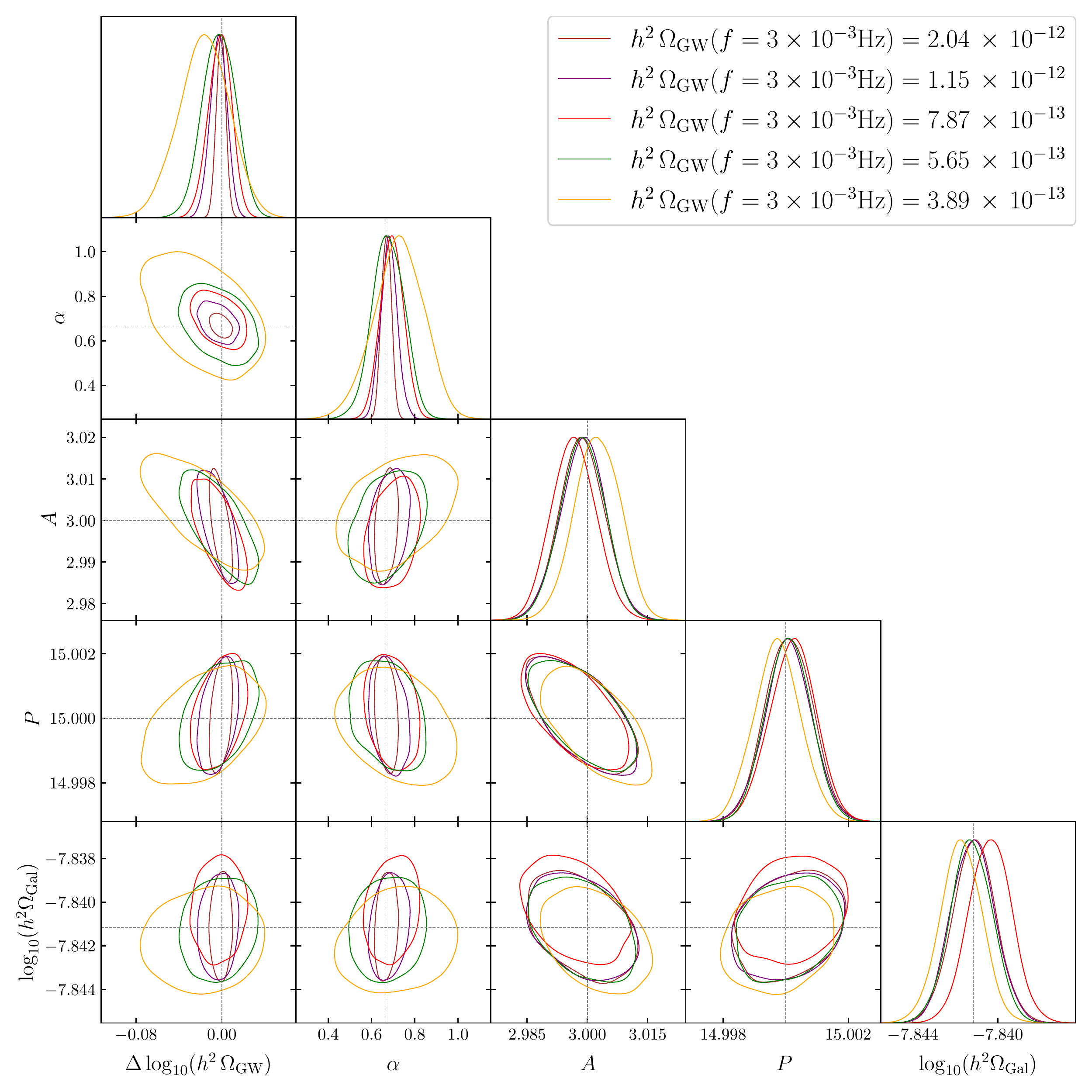}
  \caption{$2$-$\sigma$ MC contours on the SOBBH SGWB parameters $(\omhsqpivot, \alpha)$, GB parameter $h^2\Omega_\mathrm{Gal}$, and instrumental noise parameters $(A, P)$ for the five percentile levels of the SOBBH SGWB plotted in \Cref{fig:bkgperc}. 
   }
  \label{fig:bkgpost}
\end{figure}

In \Cref{fig:bkgpost} we show the MC contours (2-$\sigma$ contours) on the SOBBH signal parameters $(\log_{10}(h^2\Omega_\mathrm{GW}(f_*)), \alpha)$, together with the parameters of the GB foreground and the noise  $(\log_{10}(h^2\Omega_\mathrm{Gal}), A, P)$, obtained by injecting each of the SOBBH SGWB posterior percentiles shown in \Cref{fig:bkgperc}.
For all the injected SGWB amplitudes, the reconstruction of both the signals and the noise is accurate, with all parameters consistent with the injected values at 2-$\sigma$. 
In particular, the simultaneous reconstruction of the GB and SOBBH SGWB is achievable even when the amplitude of the latter is  small, due to their different spectral shapes.

The uncertainty on the SOBBH SGWB amplitude from the LISA measurement is practically constant: $\Delta \omhsqpivot\approx 4\times10^{-14}$ ($68\%$ CI), with relative uncertainties ranging from $5\%$ for $P_5$, to $1\%$ for $P_{95}$.
Consequently, the $< 1\%$ computation error on the SGWB prediction due to limiting $z_\mathrm{max}=5$ 
is acceptable for this study (cf.~\Cref{fig:analyticalSignalII}). 
Moreover, LISA is insensitive to the $\approx 0.2\%$ population sample variance on the amplitude
(cf.~\Cref{fig:char_strain_realisations}).

Rather than sampling over the tilt $\alpha$, as we did in the present background-detection study, in a realistic data analysis pipeline 
searching for the SOBBH SGWB, the tilt would be fixed to $\alpha=2/3$.
Thus, LISA's determination of the background amplitude 
could reveal more accurate, with respect to the tilt-marginalised 
errors presented here. 
On the other hand, realistic data 
would contain the contribution from all the other GW sources in the LISA band, which need to be extracted simultaneously to the SGWBs signals, possibly affecting the error on the SGWB amplitude compared to the simple MC evaluation performed here (see e.g.~\cite{Littenberg:2023xpl}).

\begin{figure}[t!]
    \centering
    \includegraphics[width=\textwidth]{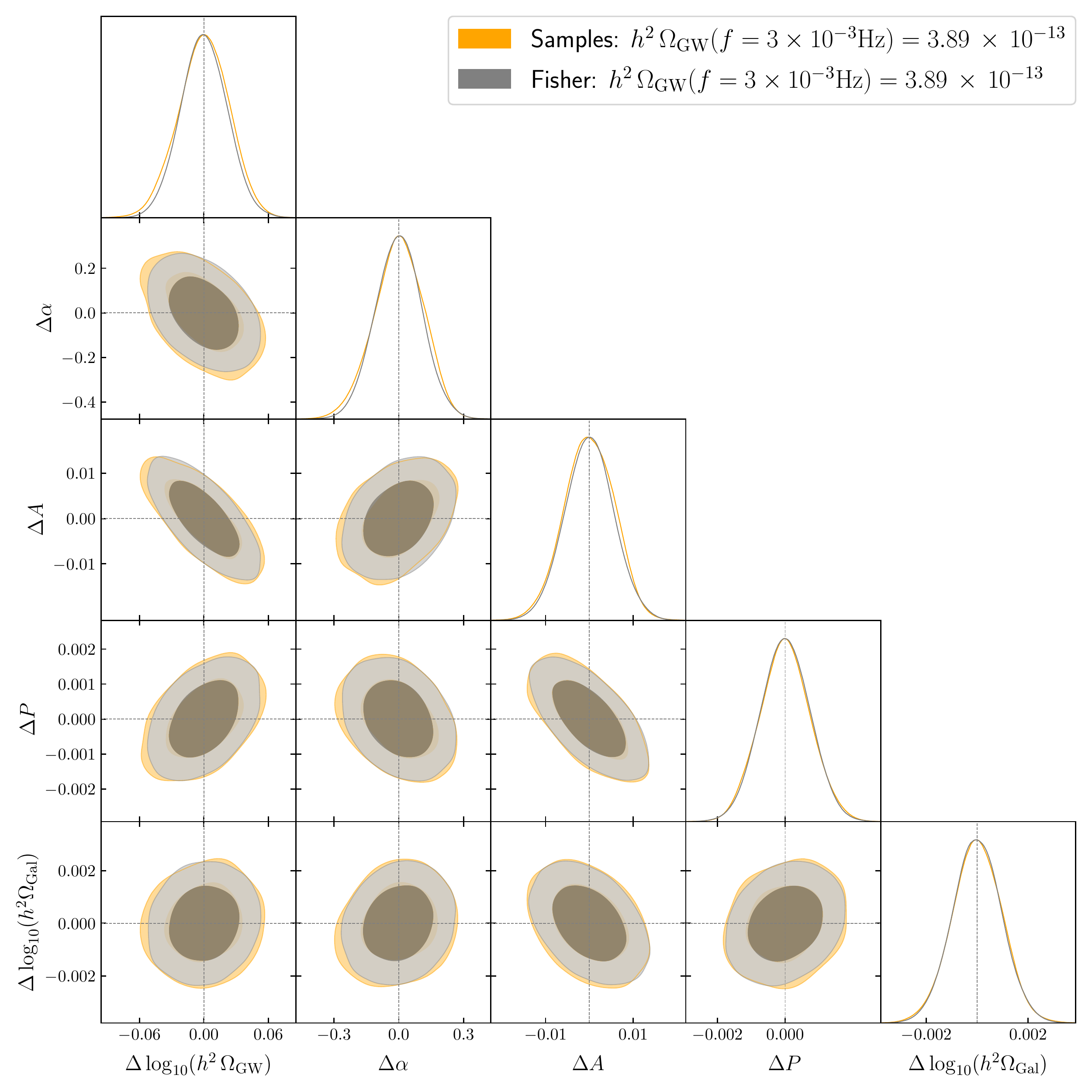}
  \caption{ 
  1 and 2-$\sigma$ contours from the Fisher analysis, compared with the MC ones, assuming the lowest value of the SGWB amplitude among the five percentile levels plotted in \Cref{fig:bkgperc}. 
  }
    \label{fig:bkglowest}
\end{figure}

As a sanity check, for the lowest value of the background amplitude, we have also performed a Fisher parameter estimation. 
In \Cref{fig:bkglowest} we present the comparison between the Fisher analysis and the corresponding MC result, showing that 
the two procedures are consistent in the reconstruction accuracy of the signal and noise parameters.

The results of this section show that LISA will be able to narrow down by one order of magnitude the current uncertainty on the  SGWB amplitude due to the SOBBH population uncertainty inferred from GWTC-3 (see \Cref{fig:bkgperc}). 
Moreover, we demonstrated that a clear detection of the SGWB is guaranteed, if the true signal falls within this uncertainty range. 
On the other hand, the lack of detection, or the detection of 
an SGWB outside the posterior 
prediction (likely lower), 
would indicate 
either that the population model needs to be changed,  
for example modifying the merger rate behaviour at high redshift, as discussed in \Cref{sec:postpredictive}, or possibly
introducing a redshift-dependence in the mass probability density function; or, it could indicate that the nature of the SOBBHs is different from what assumed in this work, for example, they could have highly eccentric orbits. 
By the time LISA will perform the SGWB measurement (or constraint), 
the SGWB amplitude posterior predicted from 
ground-based observations will probably have narrowed, if not a detection be made by either 2G or (more likely) 3G detectors.  
Nevertheless, the LISA measurement will provide further insight into the population of inspiralling SOBBH, by probing the population properties at high redshift and with low masses, and by testing the SGWB signal in a different frequency window.

\subsection{Impact
on the Power-Law Sensitivity}
\label{sec:powerlawsens}

\begin{figure}[t!]
    \centering
    \includegraphics[width=\textwidth]{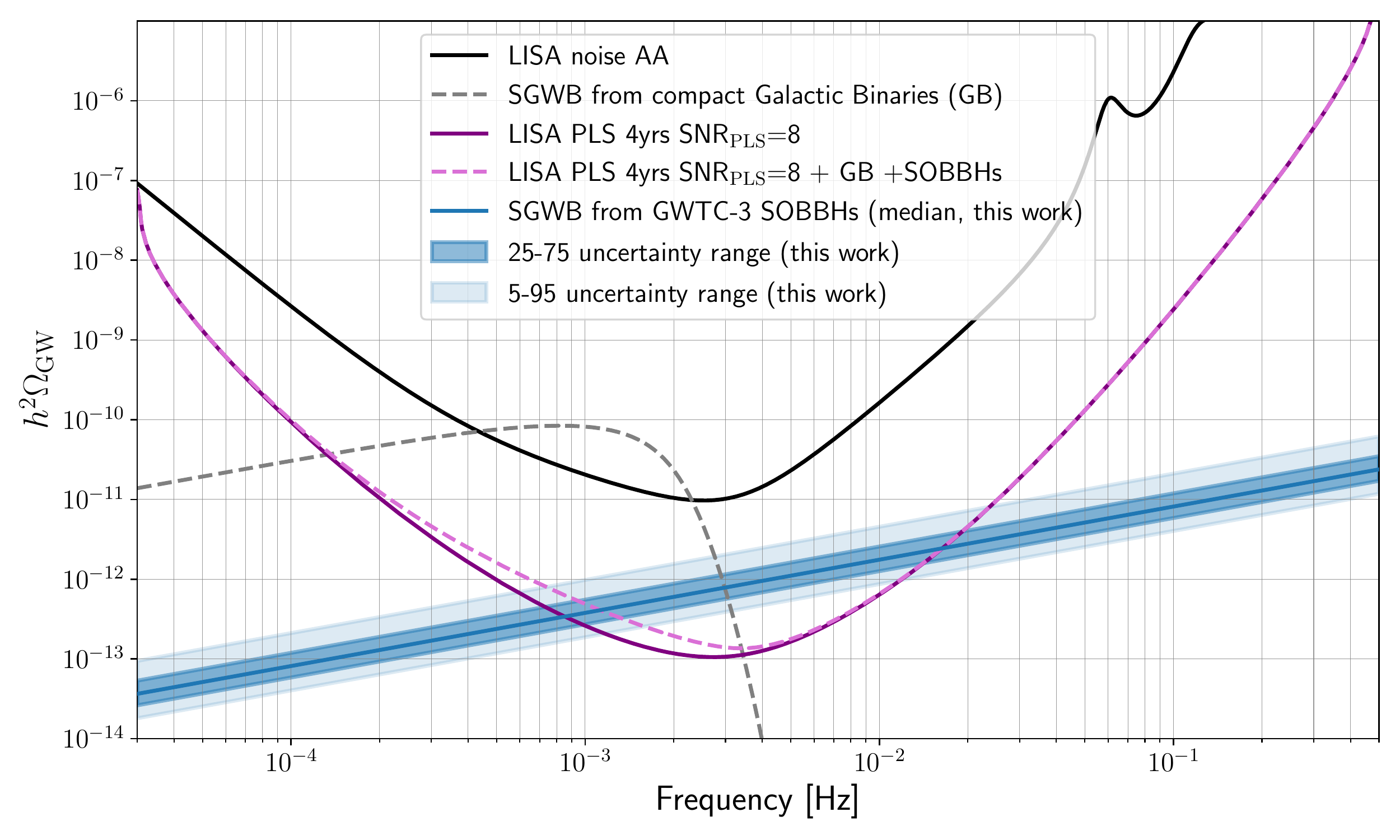}
  \caption{  Effective, i.e., averaged over all TDI channels, LISA PLS for 4 years of observations (with $100\%$ efficiency), including (in dashed purple) and excluding (solid purple) the GBs and SOBBHs SGWB components. The black line shows the sensitivity of the AA TDI channel, and the dashed grey line shows the amplitude of the SGWB due to unresolved GBs.
  The median value for the SOBBH SGWB estimated in this work from GWTC-3 constraints on the SOBBH population (with 25-75 and 5-95 uncertainty ranges) is shown in blue.
  \label{fig:PLSbkgpost} }
\end{figure}

The PLS represents the standard tool to
estimate the observability of a given power-law SGWB. 
The PLS is normally defined assuming that the only stochastic component 
affecting the SGWB measurement is the instrumental noise~\cite{Thrane:2013oya, Caprini:2019egz, Flauger:2020qyi}. 
In \Cref{fig:PLSbkgpost} we present an improved version of the LISA PLS including 
the confusion noises generated by GBs and by SOBBHs. 
For the GBs we adopt the analytical template of \cite{Karnesis:2021tsh} with all the parameters taken at their reference value; 
the SOBBH amplitude on the other hand is fixed to 
the median value evaluated in this analysis, see \Cref{sec:postpredictive}, and the tilt to $2/3$.

The GB contribution mainly affects the low-frequency range, while the SOBBH contribution is relevant at higher frequencies: this effect is reflected in the PLS.  
The inclusion of the GB confusion noise slightly modifies the PLS at low frequencies, while the impact of the SOBBHs is nearly negligible. 
Note that \Cref{fig:PLSbkgpost} corresponds to Figure 2 of \cite{Lewicki:2021kmu}, while Figure 3 in the same reference is relative to a different treatment, meant to account for the effect of the SGWB amplitude uncertainty, evaluated from the  GWTC-2 uncertainty on the merger rate at $z=0$.

\subsection{SGWB detection and the SOBBH population parameters}
\label{sec:impactpopparams}

\begin{figure}[t!]
    \centering
    \subcaptionbox{%
    \label{fig:bkgconstraints_rate}}%
    {\includegraphics[width=0.48\textwidth]{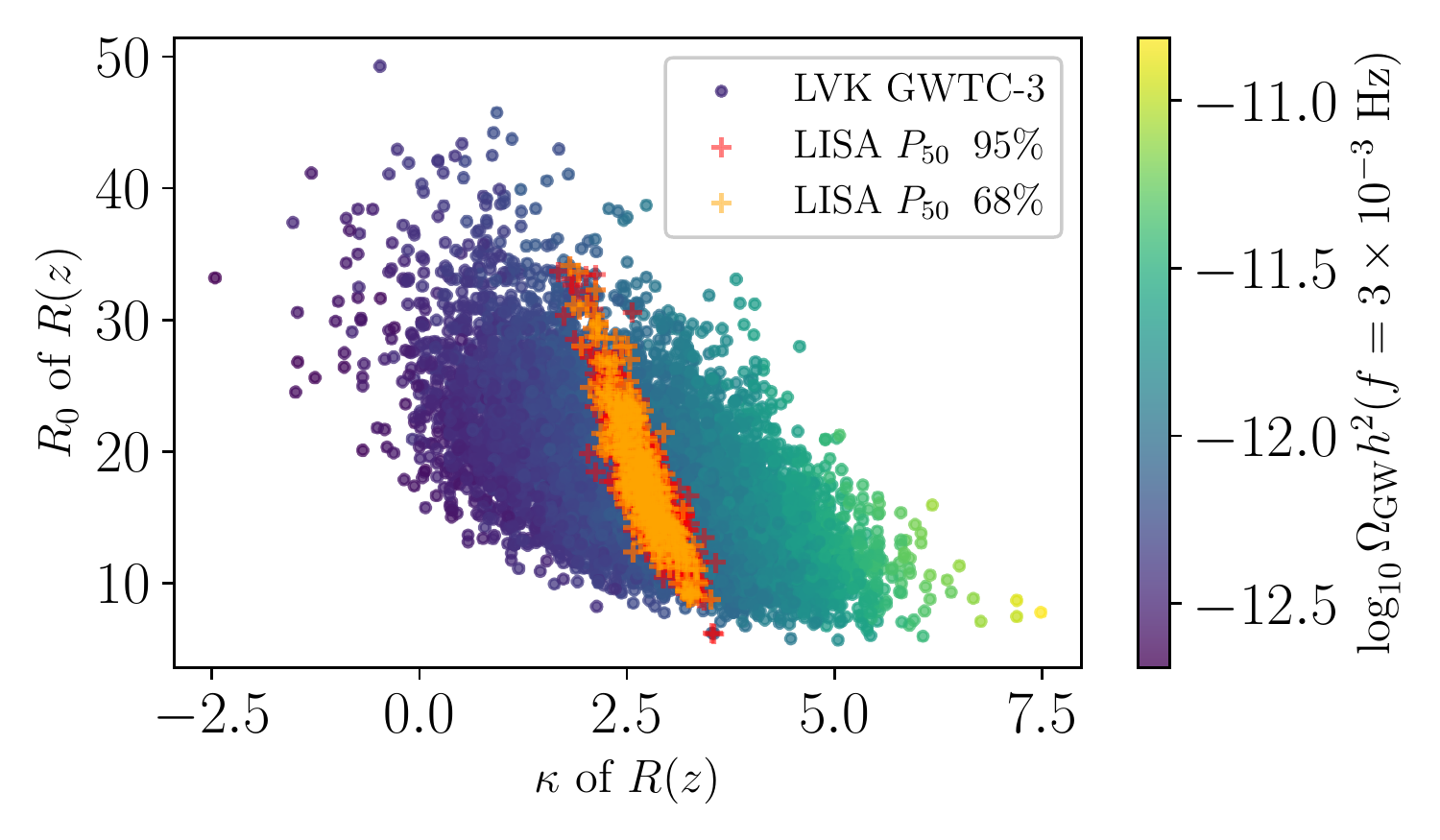}}%
    \subcaptionbox{%
    \label{fig:bkgconstraints_mass}}%
    {\includegraphics[width=0.48\textwidth]{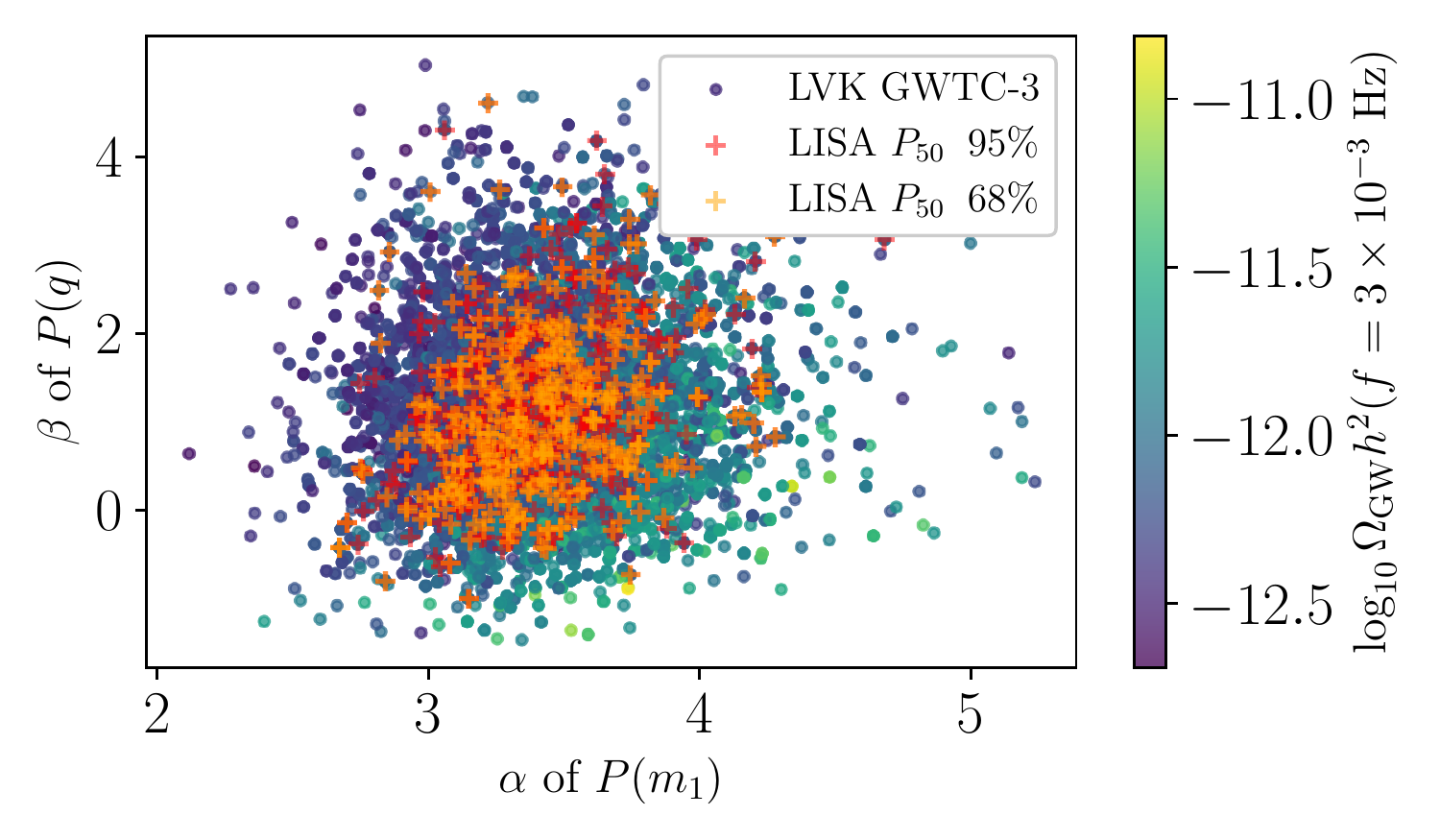}}%
  \caption{Impact, on the population parameters posterior inferred from GWTC-3, of the measurement by LISA of 
  a SGWB with amplitude corresponding to the median value $P_{50}$ of the amplitude distribution given in \Cref{fig:bkgperc}. 
  The points coloured from blue to yellow  represent the GWTC-3 posterior, and the color scale represents the corresponding SGWB amplitude. The points highlighted in yellow (red) represent the parameter values providing SGWB amplitudes within the 1-$\sigma$ (2-$\sigma$) confidence region of the LISA measurement.
  Left panel: the initial merger rate $R_0$ versus its tilt $\kappa$. Right panel: the tilts $(\alpha, \beta)$ of the power-law distributions of $m_1$ and $q=m_2/m_1$ respectively. 
  A measurement of the SOBBH SGWB would break the degeneracy coming from constraints based on individual mergers, and the credible intervals would shrink correspondingly, especially for the merger rate parameters. 
  Had we not fixed the high-redshift behaviour of the merger rate, but treated it probabilistically, the improvement with respect to the GWTC-3 constraints would be smaller, but still significant.}
    \label{fig:bkgconstraints}
\end{figure}

Intuitively, one might expect the constraining power of a measurement of the SGWB on the SOBBH population model to be very limited, regardless of its precision, since it would reduce the dimensionality of the population parameter space at most by one, leading to a highly-degenerate posterior. 
On the other hand, this can still have an important impact if the degeneracy associated with the SGWB measurement does not align with the correlations in the population parameter posterior associated with the detection of individual events, the misalignment being due to the fact that the population parameters influence the SGWB amplitude differently from how they influence the characteristics of the population of individually resolvable events. 
Indeed, it has been demonstrated that a SGWB measurement (or even upper limit) by LVK, in combination with resolved merger events, can constrain the redshift evolution of their merger rate \cite{Callister:2020arv,KAGRA:2021kbb} and possibly their mass distribution \cite{Jenkins:2018kxc}. 

The high precision with which LISA is expected to measure the SOBBH background, as shown in \Cref{sec:backgroundparamsestimation}, should render LISA especially suited to this task.
In order to illustrate its potential constraining power, in \Cref{fig:bkgconstraints} we plot the GWTC-3 population parameters posterior sample as a scatter plot, highlighting the points compatible with a SGWB amplitude within the LISA 1- and 2-$\sigma$ credible intervals, relative to a detection by LISA of a SGWB with amplitude corresponding to the median predicted SGWB level $P_{50}$ (see \Cref{fig:bkgpost}). One can appreciate that the two-dimensional posterior shrinks significantly, depending on the combination of population parameters.

In the left panel of \Cref{fig:bkgconstraints} we show the local merger rate $R_0$ versus its low-redshift tilt $\kappa$: the GWTC-3 posterior (points coloured in blue to yellow (for increasing SOBBH SGWB amplitude) presents a degeneracy due to the merger rate being best determined around $z\approx0.2$. 
Since the value of the low-redshift tilt $\kappa$ has a strong impact on the SGWB amplitude, the latter varies considerably along this degeneracy (colour scale from blue to yellow). 
Thus, a precise SGWB measurement, as performed by LISA, would break this degeneracy by leading to a posterior, in the $(R_0,\kappa)$ parameter plane, almost perpendicular to the one inferred from the detection of individual SOBBH merger events by ground-based observatories.

The posterior distribution of the mass tilts $(\alpha, \beta)$, shown in the right panel, would also be significantly reduced.\footnote{
\Cref{fig:analyticalSignal} highlighted the effect on the SGWB amplitude of $(\lambda,\sigma)$, the parameters of the peak in the \PP $m_1$ distribution, rather than the one of $(\alpha,\beta)$, the tilts of the power laws in  $m_1$ and $q=m_2/m_1$. 
The analysis performed in this section shows the converse, hence the choice of parameters in the right panel of \Cref{fig:bkgconstraints}, apparently contradicting \Cref{fig:analyticalSignal}.
However, in \Cref{sec:analytic} the impact of the population parameters uncertainty on the SGWB amplitude was considered for each parameter individually so that the influence of the posterior degeneracies was disregarded.
} 
Note that this could be further exploited by a measurement of the anisotropic component of the SOBBH background \cite{Bartolo_2022}, since the relative amplitude of the anisotropic to the isotropic components appears to be correlated with the tilt of the mass distribution and with the maximal allowed mass \cite{Jenkins:2018kxc}.

The above results are valid within the assumptions of our analysis, in particular, that the merger rate at high redshift is fixed to the SFR as given in \Cref{eq:rz}, and that the LISA uncertainty on the SGWB amplitude is inferred from the MC analysis of a simulated data set containing exclusively the SOBBH SGWB, the GBs, and the instrumental noise. 
Allowing for variations in the high-redshift model of the merger rate, and/or performing a more realistic data analysis procedure accounting for the overlap of several categories of LISA sources, would likely reduce the potential of the SGWB measurement to shrink the population parameter posterior. 
However, these effects are not expected to alter the misalignment of the correlations in the posterior parameter space inferred from the  measurements of individual events and from the measurement of the SGWB.
Consequently, the latter will anyway retain, to some degree, its constraining power.

\section{Conclusions}
\label{sec:conclusion}

We have evaluated the SGWB expected in the LISA frequency band from SOBBHs, incorporating the most recent information on their mass function, spin distribution, and merger rate coming from LVK observations, in particular from the GWTC-3 posterior on the population parameters of the \FID model. 

The LVK observations only probe the SOBBH population at low redshift, while faint and distant SOBBHs contribute the most to the background signal. 
In order to properly evaluate the SGWB, we have therefore extended the GWTC-3 power-law merger rate by assuming that it follows the SFR \cite{Madau:2016jbv}, since the low-redshift expansion of the latter is coherent with the GWTC-3 constraints.
With the aim of assessing the impact that this assumption has on the SGWB amplitude, we have also added a time delay in the SOBBHs formation and found that (under a simple model for the time delay distribution, and reasonable values for the minimal time delay) this reduces the SGWB amplitude by at most 40\%, remaining within the uncertainty inherent to the GWTC-3 posterior. 
Though the current precision of the model is not sufficient, future ground-based observations of individual merger events, together with a
detection of the SGWB by LISA, will allow to constrain the merger rate and possibly
time delays in the future.

We have used four methods to estimate the SGWB. 
The first method is based on analytic considerations and consists of the integration, over the number density of binaries, of their GW emission in the quasi-circular Newtonian approximation, resulting in a power-law SGWB with slope $f^{2/3}$ \cite{Phinney:2001di}. 
The analytical approach has been used to evaluate the impact of each population parameter on the amplitude of the SOBBH background, accounting for its marginalized 95\% confidence levels from the GWTC-3 posterior.
The power-law index $\kappa$ of the low-redshift expansion of the merger rate  is the parameter influencing the most the SGWB amplitude. 
We have then calculated the relative percentage change induced in the latter by varying the redshift upper cutoff:
we found that integrating up to $z_\mathrm{max} = 5$ is sufficient to obtain $\sim 1\%$ accuracy in the evaluation of the background amplitude, also well within the uncertainty induced by the GWTC-3 posterior. 

The other three methods employed for the SGWB estimation, gradually increasing in complexity and accuracy, rely on synthetic SOBBH populations, which we have constructed following the GWTC-3 \FID posterior distribution. 
The second method consists in replacing the integration of the analytical method with a MC sum over the masses and redshift of the SOBBHs in the synthetic population realisation, averaging over the time-to-coalescence and the sky-position; while in the third method, the MC sum is performed accounting for the time-to-coalescence of individual events and binning them according to their corresponding emission frequency. 
These methods allow establishing that the impact of the population variance over the SGWB amplitude is of the order of 0.2\%, negligible with respect to the effect of the maximal redshift choice, which is in turn smaller than the uncertainty due to the GWTC-3 posterior.  

The fourth method incorporates the actual LISA detection process and consists in simulating LISA data-streams containing the waveforms of all the SOBBHs within the simulated population, and iteratively subtracting the loudest ones until only the confusion noise remains \cite{Karnesis:2021tsh}. 
The threshold SNR used to single out the resolvable GW sources is set to $\rho_0=8$, but we find that the saturation threshold, above which the SOBBH signal is less sensitive to the choice of the threshold itself, is situated at $\rho_0\approx 5$. 

We have checked that the four methods provide consistent results for the SGWB amplitude: this is indeed the case at frequencies higher than about 2 mHz, this threshold being exclusively due to the computational limitation of our synthetic populations. 
The SNR threshold choice results in fact in a limited number of resolvable events, so that the SGWB in the LISA frequency band does not deviate from the analytical power law prediction, which is reproduced also by the three methods based on population synthesis. 
However, if sources with SNR lower than five will be resolvable in the future, thanks to improvements in data analysis techniques, or to archival searches using future ground-based detector observations, it will be necessary to take into account that the shape of the SGWB in the LISA band deviates from the power law behaviour.  
This clearly stresses the importance of a precise identification of the resolved sources and of their subtraction, which we present in a follow-up paper \cite{next}. 

The distribution of the SGWB amplitude at the reference frequency of 3 mHz is evaluated using the analytical method, for all points in the GWTC-3 posterior parameter sample.
The interquartile range of the distribution is $\omhsqpivot \in [5.65,\,11.5]\times10^{-13}$.
Our findings are in broad agreement with previous evaluations of the SOBBH stochastic signal and appear therefore to be robust with respect to assumptions such as the high-redshift behaviour of the merger rate and the mass distribution.

We have then performed a MC analysis of simulated LISA data to infer the parameters (i.e., amplitude
and spectral tilt) of the SOBBH SGWB in the presence of instrumental noise
and of the stochastic signal from GBs. 
We have found that, with four years of data, the template-based reconstruction of the parameters of both signals and of the noise is accurate to the percent level, with all parameters consistent with their injected values at 2-$\sigma$. 
In this simplified setting where no other GW source is present, and the GB background is static, the SOBBH SGWB can therefore be distinguished from the GB one, despite their overlap at low frequency.  
We have also compared the MC analysis with a Fisher Information Matrix analysis, finding good agreement, and derived the PLS accounting for the SOBBH and GB backgrounds.

The precision with which LISA will measure the amplitude of the SOBBH SGWB goes from at best 1\% (at 1-$\sigma$), for the amplitude value corresponding to the 95th percentile of its posterior distribution, up to 5\% for the fifth percentile. 
LISA will therefore reduce by one order of magnitude the current uncertainty on the SGWB amplitude predicted from the GWTC-3 population model.
The accuracy of this measurement opens interesting perspectives. 
We have shown that LISA has the potential to break the degeneracy between some population parameters, since the correlations in the posterior parameter space inferred from the
measurements of individual events and of the SGWB, are almost orthogonal.  
In particular, we forecast an important impact on the merger rate parameters, since the SGWB detection by LISA probes the population of inspiralling SOBBHs at high redshift, fully complementary to actual ground-based observations of low-redshift mergers.

Several extensions of our work are possible, tackling some of its underlying assumptions. 
First of all, including eccentricity and precession in the waveforms might have an important effect on the SGWB \cite{DOrazio:2018jnv,Zhao:2020iew}.
While we have shown 
the effect of introducing a time delay between the star formation and the BHs mergers, the impact of the metallicity on the BH mass function has been neglected, see e.g., \cite{Dvorkin:2016wac,Santoliquido_2020,Fishbach_2021,van_Son_2022} for recent studies. 
A further layer of complexity can be added including the possibility of a redshift dependence of the mass function~\cite{LIGOScientific:2021psn}.
The LISA error on the SGWB parameters should be forecasted including other types of sources in the data stream, both resolved and of stochastic nature. 
Extra-galactic neutron star binaries, for example, generate a collective signal that, although lower in amplitude, is similar to the SOBBHs one, and likely not negligible. 
Extreme mass ratio inspirals \cite{2007CQGra..24R.113A} also produce a background at mHz frequencies, although its amplitude is currently poorly constrained and its frequency dependence might not follow a simple power-law in the LISA band \cite{2020PhRvD.102j3023B}.
Finally, the effect of the SGWB measurement by LISA on the SOBBH population parameters demonstrated in this work should be properly evaluated via a joint analysis of simulated data from LISA and ground-based observatories, possibly 3G detectors which might be operative by the time LISA flies \cite{Branchesi:2023mws}. Such a joint analysis may also reveal deviations from the expected SOBBH SGWB spectrum, which could point towards a different origin for the BBHs (see e.g.\ \cite{Braglia:2021wwa}).

\acknowledgments
The authors thank Léonard Lehoucq and Irina Dvorkin for useful discussions.
C.C.\ thanks M.\ Vallisneri for enlightening discussions. 
D.G.F.\ thanks Androniki Dimitriou for tips on the numerical integration of the analytic evaluation of the SGWB.
M.P.\ and A.R.\ thank E.\ Barausse for useful discussions on related topics. 
A.S.\ acknowledges the financial support provided under the European Union’s H2020 ERC Consolidator Grant ``Binary Massive Black Hole Astrophysics'' (B Massive, Grant Agreement: 818691). D.G.F.\ (ORCID 0000-0002-4005-8915) is supported by a Ram\'on y Cajal contract with Ref.~RYC-2017-23493. This work was supported by Generalitat Valenciana grant PROMETEO/2021/083, and by Spanish Ministerio de Ciencia e Innovacion grant PID2020-113644GB-I00.
G.N.\ is partly supported by the grant Project.~No.~302640 funded by the Research Council of Norway.
A.R.\ and J.T.\ acknowledge financial support from the Supporting TAlent in ReSearch@University of Padova (STARS@UNIPD) for the project ``Constraining Cosmology and Astrophysics with Gravitational Waves, Cosmic Microwave Background and Large-Scale Structure cross-correlations''.

\appendix

\section{Further information on the SOBBH population model}
\label{app:popmodel}

In this Appendix, we provide more detail on the SOBBH population model: we describe the characteristics of the probability distributions inferred from GWTC-3 observations \cite{LIGOScientific:2021psn}, and justify some of our choices for the catalogues generation, in particular regarding the merger rate behaviour with redshift and the maximal time-to-coalescence.

\subsection{Redshift-dependent SOBBH rate}
\label{app:mergerrate}

As discussed in \Cref{sec:postparams}, the GWTC-3 constraints on the SOBBH merger rate variation with redshift, assumed to be a power law $R(z)=R(0)(1 + z)^\kappa$,  are weak at $z\gtrsim 0.5$. 
Therefore, in order to produce accurate SGWB estimates, we need an Ansatz that extends the power law assumption towards higher redshift. 
We require $R(z)$ to follow the redshift profile of the Madau-Fragos SFR \cite{Madau:2016jbv}:
\begin{equation}\tag{\ref{eq:rz}}
    R(z) = R_{0} C \frac{(1 + z)^\kappa}{1 + \frac{\kappa}{r}\left(\frac{1 + z}{1 + z_\text{peak}}\right)^{\kappa + r}}\,,
\end{equation}
with $r,\kappa>0$ and $R_0 \equiv R(z=0)$, implying  $C = 1 + (\kappa/r) \left(1 + z_\text{peak}\right)^{-(\kappa + r)}$. 
Thus, along the evolution of the universe, from high to low redshift, the SOBBH merger rate initially \emph{rises} as $z^{-r}$ as more stars are available, and eventually \emph{decreases} as $z^{\kappa}$ after the peak of stellar formation. 
Different from previous studies, e.g.\ Ref.\ \cite{KAGRA:2021kbb}, we introduce the extra factor $\kappa/r$ in the denominator of \Cref{eq:rz} to guarantee that the function peaks precisely at redshift $z_\text{peak}$; otherwise, the actual peak of the function would deviate from the value of the nominal $z_\mathrm{peak}$ parameter whenever $\kappa/r\neq 1$. 
Following this notation, the updated best fit values found in \cite{Madau:2016jbv} are $\kappa=2.6$, $r=3.6$, and $z_\mathrm{peak}=2.04$.

\begin{figure}[!t]
  \centering
  \includegraphics[width=0.8\columnwidth]{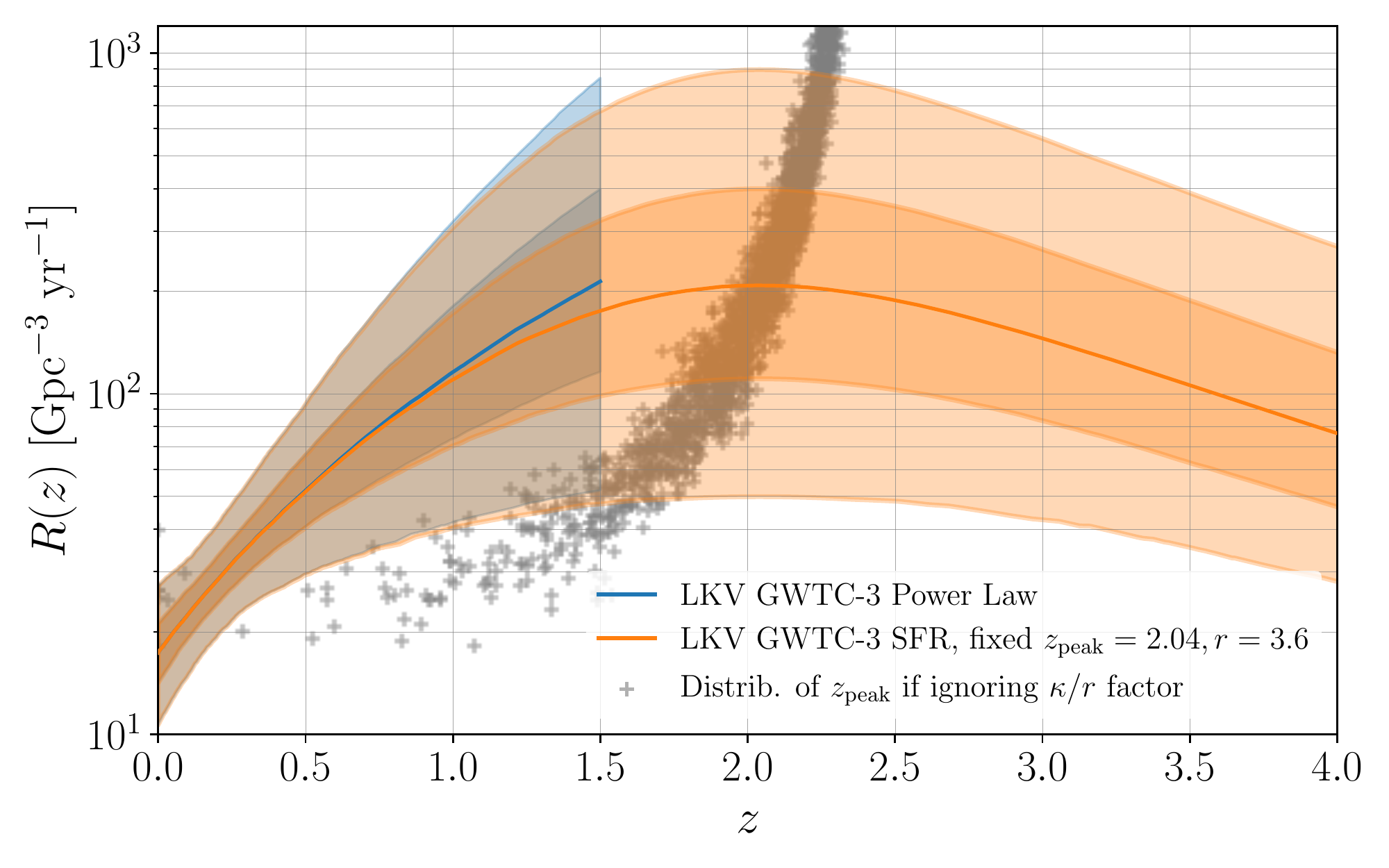}
  \caption{Distribution (median, $25$--$75$ and $5$--$95$ percentile ranges, with different levels of opacity) of the power-law merger rate $R(z) \propto (1+z)^\kappa$ (blue) according to the probability density obtained in \cite{LIGOScientific:2021psn} for the \FID model; and its extension to high-$z$ beyond the low-redshift power-law behaviour, according to \Cref{eq:rz} (orange), with fiducial values $z_\mathrm{peak}=2.04$, $r=3.6$. 
  By construction, the two distributions are very similar in the redshift range probed by LVK, i.e.~$z\lesssim 0.5$.  The crosses (gray) represent a sample of positions of the ill-defined peak of $R(z)$, had we ignored the $\kappa/r$ factor in \eqref{eq:rz}.
}
\label{fig:merger_rate_comp}
\end{figure}

In order for the merger rate $R(z)$ of \Cref{eq:rz} to work as a reasonable high-redshift extension of the GWTC-3 low-redshift constraints, we require it to reproduce the profile that LVK obtains for the \FID fiducial model fitting the GWTC-3 data~\cite{LIGOScientific:2021psn}. In that study, inference is performed on a low-redshift power law $R(z)\propto(1+z)^\kappa$, resulting in\footnote{All parameter ranges are given as median $\pm$ its respective differences with the percentiles $5$ and $95$, taken from the public population posterior sample of GTWC-3 for the fiducial \FID model.} $\kappa=\pmint{2.7}{1.9}{1.8}$ and a pivot rate of $R_0=\pmint{17.3}{6.7}{10.3}\,\Runits$ at $z=0$, or alternatively $R_{0.2}=\pmint{28.3}{9.0}{12.9}\,\Runits$ at $z=0.2$. These constraints are represented by the blue-shaded region in \Cref{fig:merger_rate_comp}. 
At low redshift, the median value for the spectral index $\kappa$ coincides with that of the SFR \cite{Madau:2016jbv}: in order to extend $R(z)$ at high redshift, we can therefore match the LVK posterior values for $R_0, \kappa$ with some values for $r, z_\text{peak}$. 
The latter could e.g.~be drawn from some prior distribution; 
for the purposes of this paper (comparing LISA's sensitivity to SOBBH SGWB amplitudes approximately compatible with the GWTC-3 population inference), it is enough to fix $r, z_\text{peak}$ to the SFR best fit values mentioned above \cite{Madau:2016jbv}. 
The resulting, GWTC-3-compatible, high-redshift merger rate is displayed in \Cref{fig:merger_rate_comp} in orange shading.

\subsection{Masses and spins density distributions}
\label{app:denfun}

The probability distribution $p(\xi | \theta)$ of \Cref{eq:master} is taken from Refs.~\cite{LIGOScientific:2018jsj,LIGOScientific:2020kqk}.  
In the fiducial \FID model, $p(\xi | \theta)$ is a separable probability density function, which can be split into a joint density function for the masses $m_{i=1,2}$ of the binary, and independent density functions for the spin amplitudes $a_i$ and tilts $t_i$:
\begin{equation}
\label{eq:ProbabilitiesFunction}
\begin{split}
    p(\xi | \theta) = &\; p(m_1,m_2| \mmin, \mmax, \deltam, \alpha, \lambda_\mathrm{peak}, \mu_m, \sigma_m, \beta_q) \\
    & \times p(a_1|\alpha_a,\beta_a) \times  p(a_2|\alpha_a,\beta_a)
      \times p(\cos(t_1),\cos(t_2)|\sigma_t,\zeta)\,.
\end{split}
\end{equation}
The separability of this distribution facilitates population synthesis since the parameters in the different components can be simulated independently (e.g.\ using \emph{inverse transform sampling} in the single-parameter cases).\footnote{The data in GWTC-3 suggest some correlations that would break this separability, such as that between high spin and mass asymmetry. For the level of the analysis in this paper, it is safe to ignore this finding.}


The mass density function is usually given in terms of the mass of the heaviest binary, by convention $m_1$, and the mass ratio $q=m_2/m_1 \le 1$:
\begin{equation}
\label{eq:MassDistribution}
p{(m_1,m2)} =
\pi_1(m_1| \mmin, \mmax, \deltam, \alpha, \lambda_\mathrm{peak}, \mu_m, \sigma_m)
\times \pi_2(q|m_1, \mmin, \deltam, \beta_q)\,,
\end{equation}
where $\pi_1$ is 
a mixture density function, times a low-mass smoothing:
\begin{equation}
\label{eq:MassDistribution1}
\begin{split}
  &\pi_1(m_1| \mmin, \mmax, \deltam, \alpha, \lambda_\mathrm{peak}, \mu_m, \sigma_m) \\
  &\quad = C_1 \left[ \left(1 - \lambda_\mathrm{peak}\right)\mathfrak{P}_{[\mmin, \mmax]}(m_1|-\alpha, \mmin, \mmax) +
    \lambda_\mathrm{peak} G_{[\mmin, \mmax]}(m_1|\mu_m,\sigma_m) \right]\\
  &\quad\quad\times S(m_1|\mmin, \deltam) \,.
\end{split}
\end{equation}
Here $\mathfrak{P}_{[\mmin, \mmax]}$ is a truncated power-law distribution with negative spectral index $-\alpha$, normalized within the $[\mmin,\mmax]$ range, $G_{[\mmin, \mmax]}$ is a similarly-truncated Gaussian density function representing a possible mass pile-up of BBHs before the SN pair-instability gap \cite{Talbot_2017}, $C_1$ is an overall normalization factor (made necessary by the presence of the smoothing function), and $S$ is a smooth cutoff for low masses that interpolates between $0$ and $1$ in the interval $[\mmin, \mmin + \deltam]$ as
\begin{equation}
\label{eq:MassSmooth}
S(m|\mmin, \deltam) =
\begin{cases} 0 & \mbox{if } m < \mmin \\
[f(m - \mmin, \deltam) + 1]^{-1} & \mbox{if } m \in [\mmin, \mmin + \deltam] \,,\\
1 & \mbox{if } m > \mmin + \deltam
\end{cases}
\end{equation}
with
\begin{equation}
\label{eq:MassSmooth2}
f(m - \mmin, \deltam) = \exp{\left( \frac{\deltam}{m - \mmin} + \frac{\deltam}
{m - \mmin - \deltam} \right)} \,.
\end{equation}
The probability density function for the mass ratio $\pi_2$ in \Cref{eq:MassDistribution} is
\begin{equation}
\label{eq:MassDistribution2}
\pi_2(q|m_1, \mmin, \deltam, \beta_q) = C_q q^{\beta_q}S(q m_1|\mmin, \deltam)  \,,
\end{equation}
where $C_{q}(m_1, \mmin, \deltam, \beta_q)$ is a normalization factor. The fact that $C_q$ depends on $m_1$ in particular makes the distribution in \Cref{eq:MassDistribution} non-separable. It can be computed as
\begin{equation}
C^{-1}_{q}(m_1) = \frac{1}{m_1} \int_{\mmin}^{m_1} \diff m_2\,q^{\beta_q}\,S(q m_1|\mmin, \deltam)\,.
\label{eq:M2Norm}
\end{equation}
The LVK analysis with the latest GWTC-3 data constraints the boundaries of the mass range as $\mmin=\pmint{5.1}{1.5}{0.9}\,M_\sun$, $\mmax=\pmint{86.9}{9.4}{11.4}\,M_\sun$. The parameters of the larger-mass distribution are inferred as $\alpha=\pmint{3.40}{0.49}{0.58}$, $\lambda_\mathrm{peak}=\pmint{0.039}{0.026}{0.058}$, $\mu_m=\pmint{33.7}{3.8}{2.3}\,M_\sun$, $\sigma_m=\pmint{3.6}{2.1}{4.6}\,M_\sun$, the spectral index of the mass ratio as $\beta_q=\pmint{1.1}{1.3}{1.8}$, and the single parameter of the smoothing function as $\deltam=\pmint{4.8}{3.2}{3.3}\,M_\sun$.


The probability distribution of each of the spin amplitudes $a_i$ in \Cref{eq:ProbabilitiesFunction} is a concave Beta distribution function in the amplitude interval (in natural units) $[0,1]$, with parameters $(\alpha_a,\beta_a)$ \cite{Wysocki_2019}:
\begin{equation}
\label{eq:SpinDistribution}
    p(a_i|\alpha_a,\beta_a) = \mathrm{Beta}(a_i|\alpha_a,\beta_a)\,.
\end{equation}
As in the LVK analysis, the subscript $a$ stands for ``amplitude'' and $\alpha_a$ should not be confused with the parameter $\alpha$ of the mass distribution $p(m_1,m_2)$. The arguments $\alpha_a>1$ and $\beta_a>1$ of the Beta distribution are linked to the expectation value and variance of the inferred amplitudes (i.e.\ all the inferred $a_1$ and $a_2$ joined into a single data set) via the relationships
\begin{equation}
    \begin{array}{lcl} \mathrm{E}[a] & = & \frac{\alpha_a}{\alpha_a + \beta_a}
    \,, \qquad
    \mathrm{Var}[a]  =  \frac{\alpha_a \beta_a}{(\alpha_a + \beta_a)^2 (\alpha_a + \beta_a +1)} \end{array}  \,,
\label{eq:SpinParameters}
\end{equation}
which the LVK analysis estimates to be $\mathrm{E}[a]=\pmint{0.25}{0.07}{0.09}$ and
$\mathrm{Var}[a] = \pmint{0.03}{0.01}{0.02}$.

Lastly, the distribution for the spin tilts $t_i$ are given by independent mixtures of an isotropic component and a truncated Gaussian component centered at perfect alignment \cite{Talbot_2017}:
\begin{equation}
  \label{eq:SpinTiltDistribution}
  p\left(\cos(t_1),\cos(t_2)|\sigma_t,\zeta\right) =
  (1 - \zeta)\left(\frac{1}{2}\right)^2 +
  \zeta \, G_{[-1, 1]}(\cos(t_1) | 1, \sigma_t) \, G_{[-1, 1]}(\cos(t_2) | 1, \sigma_t)\,.
\end{equation}
For the values of this distribution, GWTC-3 infers $\zeta=\pmint{0.66}{0.52}{0.31}$ and $\sigma_t=\pmint{1.5}{0.8}{2.0}$.

\subsection{Time-to-coalescence and frequency of emission}
\label{app:ttot}

Here we discuss the role of the residual time to coalescence for the population synthesis and the SGWB computation.

A correct prediction of the SOBBH SGWB in the LISA band implies catalogues complete enough to adequately simulate the signal.
On the other hand, the only observational knowledge we have on these sources comes from LVK observations, which probe the population of merging SOBBHs. 
In \Cref{sec:popmodel} we have shown that, under the hypothesis that the binaries formation, and therefore their coalescence rates, is in a steady state, the binary rate $R(z,\tau_c)$ in \Cref{eq:master} is indeed equivalent to the one of the merging binaries, constrained by LVK observations. 
This allows us to construct the catalogues and consequently the SGWB estimation based on the LVK GWTC-3 posterior. 

The hypothesis that the binary formation is in a steady state implies that we sample the time-to-coalescence of the binaries in the catalogues uniformly in the range $\tau_c^{\rm (det)} \in [0,\taumax]$. 
We have imposed $\taumax=10^4$ yrs, much smaller than the typical time over which the SOBBH population is expected to change, $\mathcal{O}(10^9)$ yrs. 
However, is this good enough to account for all the binaries emitting in the LISA band for the entire mission duration? 
In other words, are the simulated catalogues representative enough of the SOBBH population relevant for LISA? 
In what follows we demonstrate that, while not complete, our catalogues do indeed provide all the  information necessary for a good characterisation of the SOBBH SGWB, as far as our choices on the time-to-coalescence are concerned.

The time interval over which we need to integrate the merger rate in order to obtain the appropriate
number of observed events is 
\begin{equation}
    T_{\rm tot} = T_{\rm obs} + T_{\rm maxBand},
\label{eq:TotReceivTime}
\end{equation}
where $T_{\rm obs}$ denotes the total detector observation time, while $T_{\rm maxBand}$ is the maximum, over all the binaries in the universe, of the portion of each binary's lifetime (i.e.~of $\tau_c$) which is spent in the detector frequency range.
While in the case of LVK $T_{\rm maxBand}$ is less than
seconds, so that $T_{\rm tot} \simeq T_{\rm obs}$, LISA probes the SOBBH population at a different stage, when they are still far away from merging.
Inserting the minimal LISA frequency $2\cdot 10^{-5}$ Hz and the minimal mass in the catalogues $m_{\rm min}=2.5~M_\odot$ (see \Cref{sec:fixedpoint})  in the Newtonian relation for circular orbits (here expressed at the detector, so that $\mathcal{M}_z$ is the redshifted chirp mass)
\cite{Marsat:2018oam}
\begin{equation}
  \label{eq:OrbitalFrequency}
  f = \frac{1}{8\pi} \left[ \frac{1}{5} \left(\frac{G\mathcal{M}_z}{c^3}\right)^{5/3} \tau_c^{\rm (det)} \right]^{-3/8}\,,
\end{equation}
one obtains the maximal time-to-coalescence $\tau_c^{\rm (det)}\simeq 2\cdot 10^{10}$ yrs
in the worst case scenario of an equal mass binary at the minimal catalogue redshift $z_{\rm min}=10^{-5}$ (see \Cref{sec:popsynthesis}).
Therefore, in the case of LISA, $ T_{\rm tot} \simeq  T_{\rm maxBand}$, and setting $\taumax=10^4$ yrs appears inappropriate by as much as $10^6$ orders of magnitude. 

\begin{figure}[t!]
  \centering
  \includegraphics[width=0.49\textwidth]{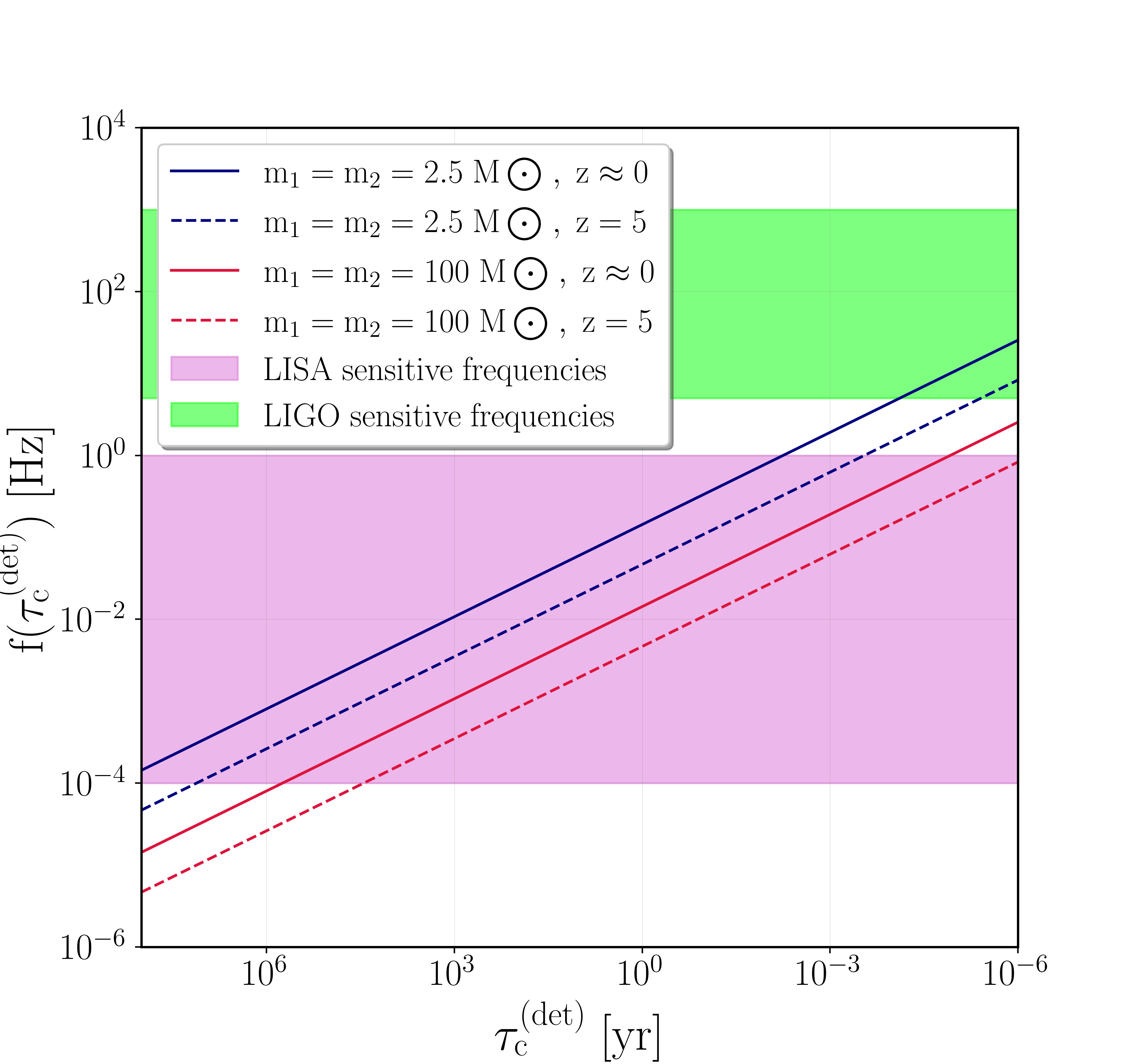}
  \includegraphics[width=0.46\textwidth]{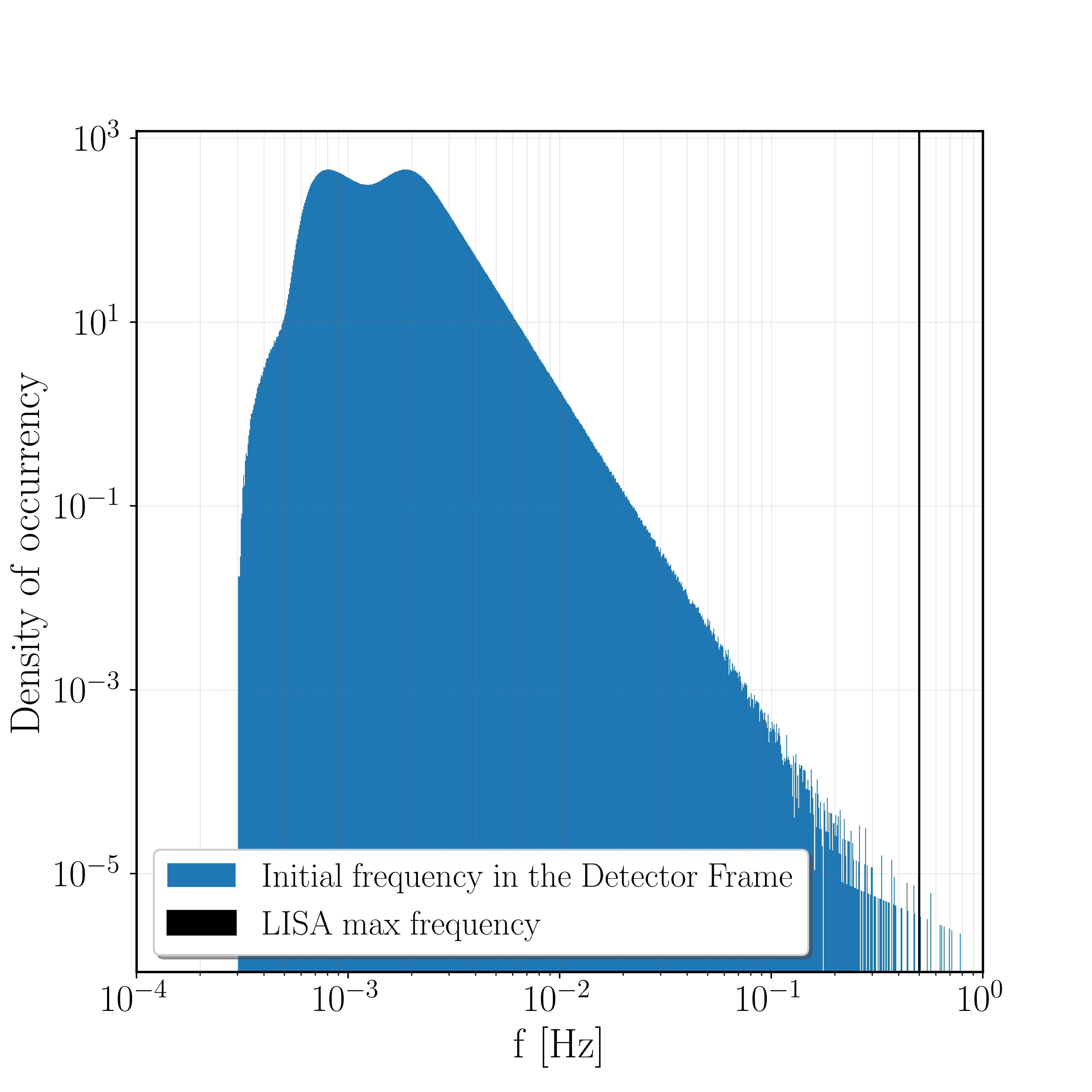}
\caption{Left panel: Evolution of the frequency during the inspiral phase as a function of the time-to-coalescence for light, mid, and large mass SOBBHs. Pink and green bands represent the LISA and LIGO frequency bands. 
Right panel: The frequency distribution emerging in one of the benchmark catalogues, constructed with a flat $\tau_c$ prior. 
}
\label{fig:InspFreq}
\end{figure}

In reality, $\taumax=10^4$ yrs is a pertinent choice that, while preserving computational feasibility, still provides all the relevant information for the SGWB evaluation.   
By cutting  the time-to-coalescence sampling at $\taumax= 5 (10) [15] \times 10^3 \,\yr$, given the catalogues mass range 
$2.5 \,M_\sun < m_2 < m_1 < 100 \,M_\sun $ (see \Cref{sec:fixedpoint}) and their redshift range $10^{-5}<z<5$ (see \Cref{sec:popsynthesis}), according to \Cref{eq:OrbitalFrequency} one is disregarding some binaries with $f\lesssim 5.9  (4.5) [3.9]\,$mHz and all binaries with $f\lesssim 0.19  (0.15) [0.13]\,$mHz, as illustrated in \Cref{fig:InspFreq}.
\Cref{fig:tceffect} shows the aggregated effect of this suppression in the SGWB (note that this figure is produced setting $z_{\rm max}=1$, as explained in \Cref{sec:fixedpoint}): it is clear from this figure that the relevant spectral property of the SGWB signal, i.e.\ the power-law behaviour in frequency, is still well captured by the signal produced via the simulated catalogues. 
The bending at low frequency is nonphysical and therefore irrelevant: the SGWB is expected to simply continue with the same power-law behaviour at low frequencies for synthetic populations with much higher $\tau_{c,\mathrm{max}}$. 
Furthermore, in \Cref{fig:PLSbkgpost}, we can see that the GB foreground overcomes the SOBBH SGWB below $2$--$3\,\mathrm{mHz}$. 
It is thus unlikely that an increase beyond $\tau_{c,\mathrm{max}}=10^4\,\yr$ would produce a noticeable effect in any realistic study. 
Given the growing computational cost of generating (and computing the SGWB of) synthetic populations with larger $\tau_{c,\mathrm{max}}$, we conclude that $\tau_{c,\mathrm{max}}^{(\mathrm{det})}=10^4\,\yr$ is a good compromise for the purposes of this study. 
%


\bibliographystyle{JHEP}
\bibliography{references}
\end{document}